\documentclass[reprint,amsmath,amssymb,aps,superscriptaddress]{revtex4-2}

\usepackage{amssymb,amsthm,bm,xcolor}
\usepackage{graphicx}
\usepackage{dcolumn}
\usepackage{bm}
\usepackage{url}
\usepackage{hyperref}
\usepackage{physics}
\hypersetup{
	colorlinks=true,
	urlcolor=purple,
	citecolor=purple,
	linkcolor=purple,
	breaklinks}
 
\usepackage{float}
\usepackage{times}

\usepackage[normalem]{ulem}

\begin{document}

\preprint{APS/123-QED}

\title{Realizing the Nishimori transition across the error threshold for constant-depth quantum circuits}

\author{Edward H. Chen}
\email{ehchen@ibm.com}
\affiliation{IBM Quantum, Almaden Research Center, San Jose, CA 95120, USA}
\affiliation{IBM Quantum, Research Triangle Park, NC 27709, USA}

\author{Guo-Yi Zhu}
\email{gzhu@uni-koeln.de}
\affiliation{Institute for Theoretical Physics, University of Cologne, Z\"ulpicher Straße 77, 50937 Cologne, Germany}

\author{Ruben Verresen}
\email{rubenverresen@g.harvard.edu}
\affiliation{Department of Physics, Harvard University, Cambridge, MA 02138, USA}

\author{Alireza Seif}
\affiliation{IBM Quantum, T. J. Watson Research Center, Yorktown Heights, NY 10598, USA}

\author{Elisa B\"aumer}
\affiliation{IBM Quantum, IBM Research, Zurich, R\"uschlikon, Switzerland.}

\author{David Layden}
\affiliation{IBM Quantum, Almaden Research Center, San Jose, CA 95120, USA}
\affiliation{IBM Quantum, T. J. Watson Research Center, Yorktown Heights, NY 10598, USA}

\author{Nathanan~Tantivasadakarn}
\affiliation{Walter Burke Institute for Theoretical Physics and Department of Physics, California Institute of Technology, Pasadena, CA 91125, USA}
\affiliation{Department of Physics, Harvard University, Cambridge, MA 02138, USA}

\author{Guanyu Zhu}
\affiliation{IBM Quantum, T. J. Watson Research Center, Yorktown Heights, NY 10598, USA}

\author{Sarah Sheldon}
\affiliation{IBM Quantum, T. J. Watson Research Center, Yorktown Heights, NY 10598, USA}

\author{Ashvin Vishwanath}
\affiliation{Department of Physics, Harvard University, Cambridge, MA 02138, USA}

\author{Simon Trebst}
\affiliation{Institute for Theoretical Physics, University of Cologne, Z\"ulpicher Straße 77, 50937 Cologne, Germany}

\author{Abhinav Kandala}
\affiliation{IBM Quantum, T. J. Watson Research Center, Yorktown Heights, NY 10598, USA}

\date{\today}
\maketitle
\textbf{
Preparing quantum states across many qubits is necessary to unlock the full potential of quantum computers.
However, a key challenge is to realize efficient preparation protocols which are stable to noise and gate imperfections. 
Here, using a measurement-based protocol on a 127 superconducting qubit device, we study the generation of the simplest long-range order---Ising order, familiar from  Greenberger-Horne-Zeilinger (GHZ) states and the repetition code---on 54 system qubits.
Our efficient implementation of the constant-depth protocol and classical decoder shows higher fidelities for GHZ states compared to size-dependent, unitary protocols.
By experimentally tuning coherent and incoherent error rates, we demonstrate stability of this decoded long-range order in two spatial dimensions, up to a critical point which corresponds to a transition belonging to the unusual Nishimori universality class.
Although in classical systems Nishimori physics requires fine-tuning multiple parameters, here it arises as a direct result of the Born rule for measurement probabilities---locking the effective temperature and disorder driving this transition.
Our study exemplifies how measurement-based state preparation can be meaningfully explored on quantum processors beyond a hundred qubits.
}

Traditionally, measurements have been synonymous with extracting information from physical systems. Yet in the quantum realm, the extraordinary nature of measurements allows them to actively modify and steer quantum states, forging a new route to entanglement generation. 
Among the more interesting entangled states are those with long-range correlations \cite{BravyiHastingsVerstraete06,Hastings2010,ChenGuWen11A,ChenGuWen11B,ZengWen15,HuangChen2015,Haah2016}; however, these cannot be prepared by any constant-depth unitary circuits, making them more sensitive to the finite coherence times of current quantum processors~\cite{mooney2021whole,mooney_generation_2021,wei2020verifying}. In contrast, recent theoretical studies have shown that the use of measurements, which are non-unitary operations, can be used to efficiently create quantum states with long-range order \cite{Raussendorf2001,Raussendorf2005,PhysRevLett.127.220503,Nat2021measure,Verresen2021cat,Bravyi2022,Lu22,zhu_nishimoris_2022,tantivasadakarn_hierarchy_2023,PhysRevLett.131.060405,Lee22,li2023symmetryenriched,buhrman2023state} and critical quasi-long-range order \cite{zhu_nishimoris_2022,Lee22,lu_mixed-state_2023}. 
In essence, measurement-based approaches trade off circuit depth for number of mid-circuit measurements and operations \cite{friedman2023locality}
as compared to exclusively unitary approaches.

In this work, we study such measurement-induced long-range order and criticality. In particular, we consider the `hydrogen atom' of long-range entangled (LRE) states, the Greenberger-Horne-Zeilinger (GHZ) state $\ket{\text{GHZ}} \propto \ket{00\cdots00} + \ket{11\cdots11}$, which can be thought of as one representative of a more general `Ising' phase of matter. 
A necessary condition for realizing GHZ is a long-range Ising order which organizes the individual qubits into a macroscopic state.
While recent experiments show the practicality of measurement-based protocols to create such Ising-like order in one-dimensional qubit geometries where stability is not guaranteed~\cite{moses_race_2023}, theoretical works suggest that this order should be robust against a range of imperfections when using a two-dimensional (2D) protocol~\cite{Preskill2002,zhu_nishimoris_2022,Lee22}. 
Here, we implement this 2D protocol on a superconducting qubit processor and, by tuning particular imperfections, we experimentally create a critical ensemble of these long-range ordered states in agreement with theoretical predictions for their stability. 

The unavoidable randomness of quantum measurements generates a `glassy' version of the sought-after long-range Ising order,
e.g.\ $\ket{00110} + \ket{11001}$, requiring some form of decoding to tame the structured randomness.
This makes it crucial to record the measurement outcomes, and then use either post-selection, feedforward, or post-processing to recover the long-range order.
In our setup, we implement post-processing to decode the hidden long-range order and determine the decoding threshold 
beyond which the order is unrecoverable \cite{Lee22} . This decoding threshold is where our quantum system exhibits a Nishimori transition, or criticality~\cite{Nishimori1981, Nishimori1993decoding},
for both incoherent \cite{Preskill2002} and coherent errors \cite{zhu_nishimoris_2022,Lee22}. 
We argue that the observed Nishimori criticality is, in fact, unavoidable in our protocol and a natural consequence 
of Born's rule -- a striking distinction from materials studies in labs seeking to observe the Nishimori criticality only by fine-tuning disorder within the material against environmental temperatures.\\

\begin{figure*}[ht]
\includegraphics[width=2\columnwidth]{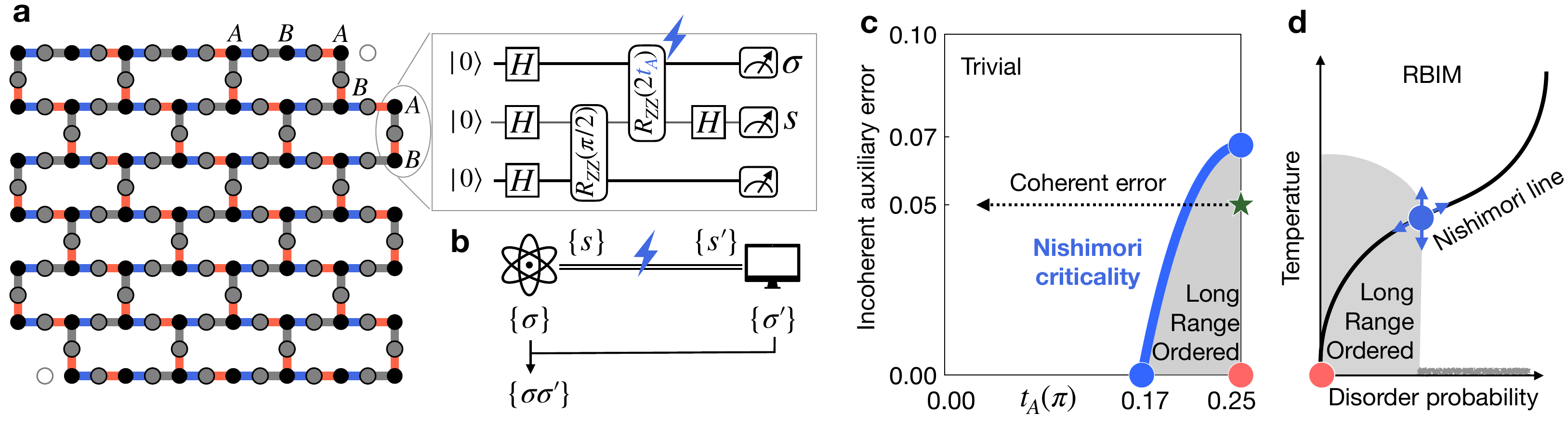}

\caption{
\textbf{
Circuit protocol, decoder, and phase diagram under coherent and incoherent errors. 
} 
\textbf{a.} The heavy-hexagonal lattice of 127 qubits. For the 125 active qubits, the inset shows the building block using constant-depth entangling circuits for the three nearest neighbors (gray circles) of each system qubit (black circles) in the presence of noise (lightning). The $R_{ZZ}$ gates are executed in order from blue, red, then gray bonds within three layers. The auxiliary outcomes, $s$, on the bonds of the lattice (gray) can be used to inform a decoder for the data outcomes, $\sigma$, on the vertices of the lattice (black).
\textbf{b.} The quantum device outputs a data bit-string $\{\sigma\}$ together with an auxiliary outcome $\{s\}$. In the presence of noise, the auxiliary outcomes become $\{s'\}$ before being passed to a classical decoder to determine a classical replica of the bit-string $\{\sigma'\}$. Their element-wise product, $\{\sigma\sigma'\}$, serves as the decoded bit-string. A measurement error (lightning) can corrupt the communication channel between the quantum replica and the classical replica. 
\textbf{c.} The trivial and long-range ordered phases sweep out distinct regions depending on the strength of coherent and incoherent noise. Within a finite threshold, a stable phase (gray), of which the GHZ is a special case (red circle), exhibits long-range entanglement in the absence of other sources of noise (e.g. without dephasing). Even in the presence of dephasing (not shown), classical long-range \textit{ordering} remains. The boundary separating the trivial and long-range phase is described by the Nishimori criticality.
Our experiments have incoherent error rates as low as $\approx$0.05, which is indicated by the green star.
\textbf{d.} Schematic phase diagram of the classical RBIM. The solid black line is the Nishimori line, which captures the {\it entire} phase diagram in \textbf{c}. 
}\label{fig:fig1}
\end{figure*}

{\bf Protocol and device operation.\ }
In our protocol, we divide the qubits on our heavy-hexagonal device into system qubits on the vertex `sites', and auxiliary qubits on the `bonds' of a honeycomb lattice (Fig.~\ref{fig:fig1}a).
We will refer to the Pauli matrices on each qubit as $X,Y,Z$.
To turn an initial product state of system qubits in $+1~X$ eigenstates into a GHZ-type state, 
we measure the $ZZ$ parities on all nearest neighbor system (site) qubits, using the auxiliary qubit in between. If the auxiliary outcome is +1, it means the two spins are perfectly anti-ferromagnetic, in the $-1$ eigenstate of $ZZ$.
A crucial element of our protocol is that we implement a coupling to the auxiliary qubit beyond a simple Clifford CNOT gate 
by an $R_{ZZ}(2t_A)=e^{-it_AZZ}$  rotation with a control parameter $2t_A$, for the $A$ sublattice (Fig.~\ref{fig:fig1}a). By varying $t_A$ away from $\pi/4$ (the Clifford limit), we can
perform tunable weak measurements or, equivalently, control the level of coherent errors.
Due to the degree-3 connectivity of the system qubits, we need to repeat the above coupling only three times before simultaneously measuring all the auxiliary qubits -- resulting in a constant-depth circuit independent of the number of qubits. 

The measurement outcomes of the auxiliary qubits in the $X$ basis, denoted by $s_{ij}=\pm 1$ for each bond $\langle ij\rangle$, are then fed as syndromes to the decoder, operated on a classical computer. The decoder produces an estimate of the quantum sample based on its limited knowledge in the form of $\{s'\}$~\cite{Preskill2002, Garratt22, Lee22,lee2023symmetry}, where $\{s'\}$ is a copy of $\{s\}$ corrupted by a finite probability $p_s$ of noise that can come from either the quantum circuit or the classical communication, as shown in Fig.~\ref{fig:fig1}b. We employ a fast decoder~\cite{Lee22} which outputs a bit-string $\{\sigma'=\pm 1\}$ as a classical ground state for each $\{s'\}$. 
By denoting the bit-string of the system qubits measured in $Z$ basis as $\{\sigma=\pm 1\}$, 
the element-wise product, $\{\sigma\sigma'\}$, between the quantum sample and the classic replica serves as the decoded bit-string. 
This is equivalent to correcting the system qubits by one layer of $X$ gates for those sites with $\sigma'=-1$, in a feed-forward manner. 

We performed experiments on \textit{ibm\_sherbrooke}, which is one of the IBM Quantum Eagle processors with 127 qubits; entangling gates generated by Echoed Cross-Resonance interactions~\cite{chow2011simple,sheldon2016ecr,malek2020ecr,sundaresan2020ecr} had typical error rates of $0.0077$ and square root of Pauli-X gates with error rates of $0.0002$ (\footnote{IBM Quantum. \href{https://quantum-computing.ibm.com/}{https://quantum-computing.ibm.com/}, 2021. (Downloaded August 6, 2023).}). 
The typical device measurement error rates of $0.010$, which were sufficiently below the decoding threshold needed for the preparation of the long-range ordered state.
%
\section*{Results}

{\bf Conceptual understanding of protocol.\ }
%
In previous theoretical work by some of the present authors \cite{zhu_nishimoris_2022}, it was shown that deviations from the Clifford limit by coherent errors induced by $t_A < \pi/4$ are tolerable up to a finite threshold.
Here we expand this perspective by also treating \textit{incoherent} errors (corrupting the syndromes)
in an analytically exact manner 
and show that the presence of both types of errors leads to a threshold line as shown in Fig.~\ref{fig:fig1}c,
which in its entirety is captured by the Nishimori criticality.
To see this, let us consider measuring the auxiliary qubits, which collapses the system's wave function into
\begin{equation}\label{eqn:isinglre}
\begin{split}
|\psi(s_{ij})\rangle = e^{-\frac{\beta}{2}\sum_{\langle ij\rangle}s_{ij}  Z_iZ_j} |+\rangle ^{\otimes N} \,,
\end{split}
\end{equation}
where $\beta = 2\tanh^{-1}\tan(t_A)$~\cite{zhu_nishimoris_2022}, and $N$ denotes the number of system qubits. 
The probability of such a measurement outcome follows from Born's rule
\begin{equation}
P(s_{ij}) = \norm{\psi(s_{ij})}^2 \propto \sum_{\sigma}e^{-\beta\sum_{\langle ij\rangle}s_{ij}  \sigma_i\sigma_j} \ ,
\label{eq:Bornrule}
\end{equation}
which resembles the partition function of the random bond Ising model (RBIM)~\cite{zhu_nishimoris_2022}. 
Concretely, by Eq.~\eqref{eq:Bornrule} we analytically map our protocol onto a RBIM precisely tracking the Nishimori line~\cite{supplement} with an effective disorder probability
\begin{equation}
	\tilde{p} = \frac{1- (1-2p_s)\sin(2t_A)}{2} \ ,
	\label{eq:disorder_probability}
\end{equation}
as a joint action of {\it both} coherent and incoherent errors that drives the phase transitions across the blue line in Fig.~\ref{fig:fig1}c. In particular, this implies that every point in the extended transition line shares the same Nishimori criticality. This scenario for the quantum protocol is quite distinct from the classical RBIM, whose schematic phase diagram is shown in Fig.~\ref{fig:fig1}d, where the Nishimori line only occurs at the fine-tuned solid line -- demonstrating an unprecedented robustness
of Nishimori criticality in the quantum case. \\

\begin{figure}[t]
\includegraphics[width=\columnwidth]{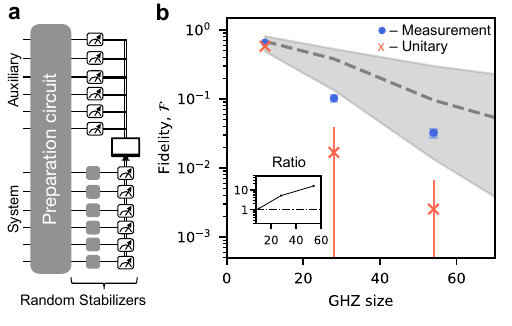}
\caption{\label{fig:fig2}
\textbf{Decoded fidelity estimation by randomly sampling GHZ stabilizers.}
\textbf{a.}
Because our decoder was implemented as Pauli corrections on the system qubits, the characterization of random stabilizers, which is measured in basis rotated by single-qubit rotations (small gray boxes), needed to be done in conjunction with the implemented decoder (symbolized by the monitor). 
\textbf{b.} Estimated fidelities relative to GHZ states for measurement-based (filled blue circles) and unitary-based (red X-marks) preparation of long-range Ising ordered states on two-dimensions. The error bars represent the standard deviation of the fidelities estimated from bootstrap resampling random sets of stabilizers (See Methods). The error bars for the unitary results (red) are comparable to the fidelity itself, thereby extending far below what is visible on a logarithmic plot; for exact values, see data. The theoretically predicted fidelities for measurement-based protocol (dashed gray line) were based on an inferred noise model with auxiliary and site readout errors with a range of parameters giving rise to a 25th-75th percentile confidence interval in shaded gray~\cite{supplement}. The inset shows the ratio of the experimentally evaluated measurement- to unitary-based fidelities increasing for system size up to 54 sites.
}
\end{figure}

\begin{figure}[b]
\includegraphics[width=\columnwidth]{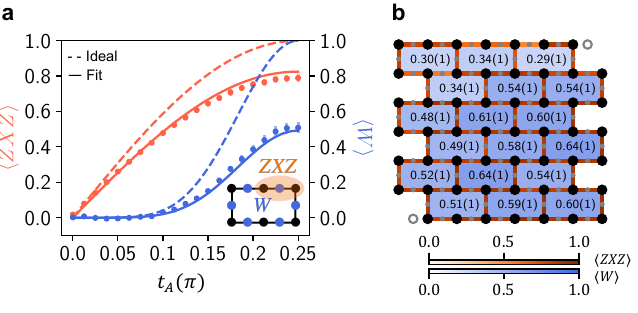}
\caption{\label{fig:fig3}
\textbf{Experimentally measured local observables used to generate the state.}
\textbf{a.} For two observables, we plot the ideally expected outcomes (dashed lines), the unprocessed experimental data (dots), and a {\it one parameter} fit (solid line) for each observable for sweeping 
$t_A$ from $0$ (trivial) to $\pi/4$ (long-range ordered). The average 3-qubit-bond (red) observable reached as high as 0.8 across the 72 total bonds, while the average 6-qubit-plaquette (blue) observable reached 0.5 across the 18 plaquettes. Although in a noiseless setting both were expected to reach unity, the measured values agree well with the fit by $p_s=5.6\%$, and $p_\sigma=2.3\%$, which are approximately consistent with the known errors on the device~\cite{supplement}. The experimental data exhibits an absence of a singularity in these observables, consistent with expectations for both local shallow quantum circuit, and the internal energy of Nishimori line. 
\textbf{b.} 125 of the 127 qubits used on \textit{ibm\_sherbrooke} where each bond ($\langle ZXZ \rangle$) and plaquette ($\langle W \rangle$) observable values are shaded according to the measured value. The numbers inside plaquettes (b) show $\langle W\rangle$ with parenthesis show standard error.
}
\end{figure}

\begin{figure*}[ht]
\includegraphics[width=\textwidth]{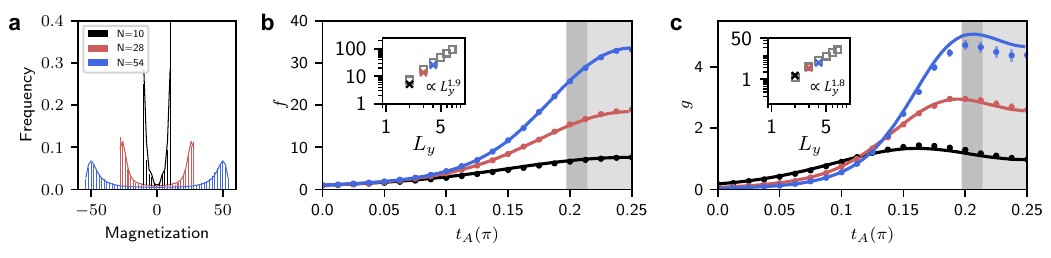}
\caption{\label{fig:fig4}
\textbf{Nishimori transition by tuning coherent errors.} 
\textbf{a.} The distribution of $M$ at $t_A=\pi/4$ for system sizes of 10 (black), 28 (red) and 54 (blue) system qubits
using 21, 63, and 125 total qubits, respectively. Solid envelope lines are theoretical estimates using a two-parameter noise model.
\textbf{b.} The sum of two-point correlation function $f$ signals the growth of long-range correlation when increasing $t_A$ and system sizes. Beyond a critical threshold for $t_A\geq t_A^c\approx 0.20\pi\sim0.21\pi$ (dark gray), the state exhibits long-range order (light gray). The estimated $t_A^c$ varied depending on the system size studied. The inset shows the size scaling of experimentally measured $f$ (X marks) at the peak location  of $g$
agreeing well with the theoretically predicted noiseless values (square markers) scaling with $\propto L_y ^ {1.9}$.
\textbf{c.} The peak locations of $g$ converge to the finite threshold that separates the long-range ordered phase (light gray) from short-range correlated phases. The dark gray shading illustrates the theoretically predicted critical point ($t_A^c$), according to the previously inferred two noise parameters, that spans a finite width because of the variation of noise probabilities. In the inset, the experimental (X marks) values agree well with the theoretically predicted values scaling as $\propto L_y ^ {1.8}$. The noiseless envelopes for all solid curves can be found in the Supplemental Materials.
}
\end{figure*}

{\bf GHZ fidelity in Clifford limit.\ }
%
For a baseline characterization of the measurement-based protocol, we estimated the fidelity of the prepared states
in the Clifford limit ($t_A=\pi/4$) relative to the GHZ state. Because the final state in this limit is a stabilizer state,
it was sufficient for a desired accuracy to consider only a constant number of randomly sampled measurements of the system qubits~\cite{Flammia11fidelity, Silva11fidelity}. For the specific case of the GHZ state, half the sampled stabilizers contain only Pauli Z operators, while the other half are combinations of Pauli-X and Pauli-Y operators~(See Methods for more details).
To assess the relative performance of our protocol, we also implemented a standard unitary protocol for constructing GHZ states~\cite{mooney_generation_2021}.
In Fig.~\ref{fig:fig2}, we see that the fidelities of the measurement-based protocol outperformed the unitary preparation. 
This can be rationalized by the latter experiencing more errors due to the long idle times of deep circuit with size-dependent depth between $O(N)$ and $O(\log(N))$.

For a system of 10 qubits, the measurement-based protocol resulted in a GHZ fidelity above 50\%, 
but with increasing system size the fidelity was found to decrease exponentially (Fig.~\ref{fig:fig2}b). 
We note, however, that this does \emph{not} imply the absence of long-range order or entanglement for these larger systems. 
In fact, we expect exponentially decaying GHZ fidelities versus system sizes in the presence of noise 
for virtually all states in the same phase of matter.
We emphasize that no form of error mitigation, for measurement or unitary gates, was used to estimate these fidelities.
To explain the experimentally measured fidelities, we compared our results against the predicted fidelities based on a noise model with $\approx 5\%$ incoherent auxiliary errors and $\approx 3\%$ data readout errors -- values inferred in the next section. 
This places us in the long-range ordered phase in Fig.~\ref{fig:fig1}c (green star), which in the absence of any additional errors, has long-range GHZ-type entanglement, whilst its predicted GHZ fidelity shown in gray in Fig.~\ref{fig:fig2} decays exponentially with the number of system qubits. We see that the experimentally obtained values are slightly suppressed with respect to the gray curve, which is likely due to dephasing. This raises the question whether we retain robust long-range order in the presence of such dephasing.  \\

{\bf Noise analysis.\ }
To determine where in the phase diagram our experimental protocol accessed the GHZ state relative to the criticality threshold -- implicitly bounding the amount of other sources of errors that were present in our experiments -- we tuned one type of coherent error, via Eq.~\eqref{eqn:isinglre}, uniformly across the device; in this sweep, we monitored and fit the experimental observables~\cite{zhu_nishimoris_2022} associated with every bond to $\langle ZXZ \rangle = (1-2p_\sigma)^2(1-2p_s) \sin(2t_A)$, and experimental observable of every plaquette to $\langle W \rangle = (1-2p_s)^6 \sin(2t_A)^6$, as shown in Fig.~\ref{fig:fig3}a (Sec.~IC in Ref.~\onlinecite{supplement}). Here $p_\sigma$ accounts for the readout error of system qubits while $p_s$ captures both readout error on the auxiliary qubits and some of the noise during the entangling process.
For $t_A=\pi/4$, the bond and plaquette observables should ideally approach unity (dashed lines) because they capture, partially, the quality of the constituent cluster states~\cite{Raussendorf2001} -- a precursor state for the GHZ state -- with experimental data shown in Fig.~\ref{fig:fig3}b. 
%
For $t_A$ below $\pi/4$, the implemented circuits become non-Clifford and thus cannot, in general, be efficiently characterized.
Nonetheless our modeling of coherent and incoherent noise sources turns out to be sufficiently comprehensive to quantitatively explain the observed experimental data, even for experiments involving up to 125 qubits. This allows us to infer the amount of noise afflicting the auxiliary ($p_s$) and system ($p_{\sigma}$) qubits when sweeping $t_A$. This led to an estimate for the amount of \textit{incoherent} errors present in the experiment to be in the range of $p_s \approx 4.2\%- 5.6\%$ and $p_\sigma \approx 1.2\%-2.3\%$ -- values consistent with our expectation based on standard calibration benchmarks of the device~\cite{supplement}. \\

{\bf Nishimori transition for tunable coherent errors.\ }
Having established the incoherent noise level of our device, we can now proceed to validate the existence of a stable, long-range, Ising ordered phase when experimentally sweeping the level of coherent errors in our protocol. 
To reveal the hidden order, we applied a decoder~\cite{Lee22, Preskill2002, PyMatching21} to process \textit{every} classical snapshot for the auxiliary qubits in the $X$ basis and the system qubits in the $Z$ basis. The basic idea is to perform a correction based solely on the auxiliary readout~\cite{zhu_nishimoris_2022}. This correction factor approximates the ground state configuration of the RBIM as a classical estimate~\cite{Garratt22}, $\{\sigma'\}$, of the bit-string from quantum device (Fig.~\ref{fig:fig1}b). 

The distribution of the decoded bitstrings in the computational basis is shown, for $t_A=\pi/4$, in Fig.~\ref{fig:fig4}a, where we sum over the {\it decoded} $Z$ expectation values of the individual qubit to obtain a total decoded `magnetization' $M = \sum_{j=1}^N \sigma_j' Z_j$. Any bias of this distribution (e.g. towards positive values) may be explained by an Ising asymmetric error originating from physical mechanisms such as amplitude damping or relaxation. Such errors would reduce the amount of classical correlations, and the small value $\langle M\rangle\approx0.02(2) N$ suggests that the global Ising symmetry is largely preserved. Moreover, the decoded, bi-modal experimental distribution (Fig.~\ref{fig:fig4}a) agrees well with the theoretical prediction (solid lines) lending confidence to the two-parameter noise model we used. 

To more rigorously characterize the long-range order, we examined, for $t_A\leq \pi/4$, the decoded system qubit bitstrings and the average two-point, \textit{classical} correlations
\begin{equation}
f := \frac{1}{N}\left(\langle M^2\rangle -\langle M\rangle^2\right) \ ,
\label{eq:QFI}
\end{equation}
which is a sum of the correlations $\langle \sigma_i \sigma_j Z_i Z_j\rangle $ for all the system qubits that compose the quantum state. 
\textbf{}The decoded experimental data is shown in Fig.~\ref{fig:fig4}b, where the solid line shows the theoretical benchmark with the noise parameters inferred from Fig.~\ref{fig:fig3}a. 
We observe a hallmark of the long-range ordered phase in the diverging $f$ for increasing system sizes; such divergent behavior for $f$ is expected throughout the ensemble of long-range ordered states, or phase, even away from $t_A=\pi/4$ up to a finite threshold, $t_A^c$. In fact, we have confirmed that in our two-parameter theory model, $f$ indeed grows unbounded above $t_A^c\approx 0.20\pi\sim0.21\pi$. In contrast, for small $t_A$ far below the threshold, $f$ is apparently bounded and does not grow with increasing size. 
This divergent behavior for our 2D protocol should be contrasted to results in 1D geometries~\cite{supplement},  
where we found $f$ to stop growing for larger system sizes in line with theoretical expectations
that $f$ is bounded by a finite correlation length in the presence of infinitesimal weak errors. 

To determine the threshold, or critical point, a practical way is to use of the normalized variance of $M^2/N$:
 \begin{equation}
 \begin{split}
g&:= \frac{1}{N^3}\left(\langle M^4\rangle - \langle M^2\rangle^2\right) \,,
 \end{split}
 \end{equation}
which quantifies the amount of fluctuations in the squared magnetization~\cite{supplement}. 
In the presence of 5\% incoherent auxiliary errors, 
the peak location is expected to converge to a critical value of $t_A^c \approx 0.205\pi$, by translating the Nishimori critical point $\tilde{p}_c\approx 6.75\%$~\cite{Queiroz2006exponents, zhu_nishimoris_2022} with Eq.~\eqref{eq:disorder_probability},
which is in very close agreement with  the experimental data where the peak locations 
approach this predicted critical point (Fig.~\ref{fig:fig4}c). 
Moreover, at this transition, we also observe that $f$ exhibits steep increases as one would expect for a critical system. 
The three experimental values for $f$ for increasing system sizes agree well with noisy classic simulations exhibiting 
a $\propto L_y^{1.9}$ scaling behavior of the peak height (Fig.~\ref{fig:fig4}b inset), where $L_y=2,3,4$ is the number of columns of qubits in a brickwall lattice; this experimentally observed scaling is in close agreement with the scaling exponent calculated value of $1.8(1)$ for the RBIM at the Nishimori point~\cite{Queiroz2006exponents}.
While the criticality is exposed in the decoded correlations only, the observable $\langle ZXZ\rangle$ of Fig.~\ref{fig:fig3}a 
is another, direct probe of Nishimori physics -- it corresponds to the internal energy of the classical RBIM along the Nishimori
line, which we experimentally confirm to be free of any singularity at the phase transition and in agreement with theoretical predictions.
\\
\begin{figure}[t]
\includegraphics[width=\columnwidth]{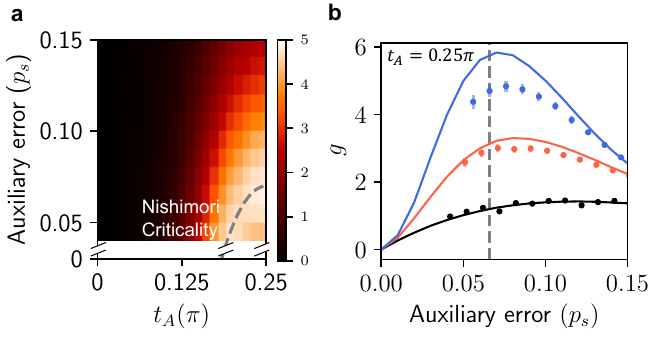}
\caption{\label{fig:fig5}
\textbf{Decoding transition out of the long-range ordered phase by increasing auxiliary errors before decoding. 
}
\textbf{a.} For the largest system size ($N=54$), we experimentally mapped the 2D phase diagram for various coherent ($t_A$) and incoherent ($p_s$) errors where the color is proportional to the amount of variance in the magnetization squared, $g$. The analytically derived contour (dashed gray) shows close agreement for incoherent, auxiliary errors starting from approximately 0.05.
\textbf{b.} For the lowest amount of injected coherent error ($t_A=\pi/4$), the experimentally estimated variance $g$ (circles) is maximized at the theoretically expected (solid lines) decoder transition of approximately $6.75\%$ (vertical dashed gray) for all three system sizes (10: black, 28: red, 54: blue).
}
\end{figure}

{\bf Decoding transition by tuning incoherent errors.\ }
%
As we have shown, the long-range ordered phase 
created by our 2D protocol is unveiled only after using a decoder, 
whose performance critically depends on the quality of the auxiliary measurements. 
While the auxiliary error is lower-bounded by the quantum device, we can inject additional errors, in post-processing, before applying the decoder (see Fig.~\ref{fig:fig1}b) and thereby chart out a broader phase diagram including varying rates of incoherent errors.
By again monitoring the degree of fluctuations, $g$, now as a function of an increasing level of incoherent errors and system size (Fig.~\ref{fig:fig5}), we observe that the Nishimori critical point $t_A^c$ shifts towards $\pi/4$ and vanishes completely at $p_s\approx 6.75\%$~\cite{Queiroz2006exponents}, the decoding threshold~\cite{supplement}. 
The origin of this limit can be readily understood as being equivalent to the decoding transition of a repetition code on a honeycomb lattice with bit-flip errors~\cite{Preskill2002}. 
Our experiments thus not only demonstrate the stability of the long-range ordered phase separated from a trivial one via a Nishimori transition, but also quantify when it would fail for more noisy devices. It also significantly distinguishes a 2D from 1D protocol where the peak quickly converges to $t_A=\pi/4$ without a finite threshold~\footnote{See Extended Data Fig.~\ref{fig:1Dv2D}}. Thus we claim that our experimentally implemented 2D protocol exhibits long-range order with intrinsic robustness.

\section*{Discussion}
%
The Nishimori multicritical point arises from a delicate balance between disorder and temperature -- a condition that is largely inaccessible in experiments on real, physical materials modeled by a RBIM~\cite{Binder86rmp}. This should be contrasted to our experiments using a shallow circuit protocol on a quantum system, where the Nishimori transition shows remarkable robustness even in a noisy device of significant size. 
We argue that this can in fact be traced back to Born's rule, which {\it naturally enforces} the delicate balance of Nishimori physics: the auxiliary qubits play the role of quenched disorder by being measured, whose {\it probability} is exactly the wave function {\it squared amplitude} of the system qubits. 

Our systematic study and generation of long-range ordered states using measurements shows that such protocols can be robust against certain errors, and even outperform unitary approaches on existing quantum hardware. Improvements in coherence and measurement fidelity should further improve the performance of our measurement-based protocol. Our work emphasizes the importance of spatial geometry in measurement-based protocols -- by \textit{tuning} errors across an error threshold we observed a stable phase that persists in 2D but is absent in 1D. While the experimentally accessible order parameters, $f$ and $g$, were observed to be below the theoretically predicted noiseless values due to the presence of noise, we expect to still be able to determine the universal critical exponents using equivalently noisy but larger devices, up to system sizes of 180, where finite-sized effects play less of a role.

It would be interesting to similarly explore the (in)stability of measurement-induced long-range entanglement upon tuning coherent and incoherent errors for other proposals in the literature \cite{Raussendorf2001,Raussendorf2005,PhysRevLett.127.220503,Nat2021measure,Verresen2021cat,Bravyi2022,Lu22,zhu_nishimoris_2022,tantivasadakarn_hierarchy_2023,PhysRevLett.131.060405,Lee22,li2023symmetryenriched,buhrman2023state,Lee22,lu_mixed-state_2023}. This is especially timely since measurements have recently been used to deterministically create exotic long-range entanglement including topological order~\cite{iqbal2023topological,fossfeig2023experimental,iqbal2023creation} and related states with dynamic quantum circuits utilizing feed-forward operations~\cite{baumer2023efficient}. In such general contexts, stability might inquire additional ingredients, such as using the time-domain~\cite{Preskill2002,Hastings2021dynamically} or higher---or even fractional---dimensions, opening up a rich territory for exploration.
\begin{acknowledgements}
We thank M. Ware, P. Jurcevic, Y. Kim, A. Eddins, H. Nayfeh, I. Lauer, G. Jones, and J. Summerour for assistance with performing experiments, and B. Mitchell, D. Zajac, J. Wootton, L. Govia, X. Wei, R. Gupta, T. Yoder, T. Soejima, K. Siva, M. Motta, Z. Minev, S. Pappalardi, S. Garratt, E. Altman, and Prof. Nishimori for thoughtful discussions. And we thank Prof. Nishimori for careful reading of the manuscript. 
The Cologne group was partially funded by the Deutsche Forschungsgemeinschaft under Germany's Excellence Strategy -- Cluster of Excellence Matter and Light for Quantum Computing (ML4Q) EXC 2004/1 -- 390534769 and within the CRC network TR 183 (Project Grant No. 277101999) as part of projects A04 and B01. The classical simulations were performed on the JUWELS cluster at the Forschungszentrum Juelich.
R.V. is supported by the Harvard Quantum Initiative Postdoctoral Fellowship in Science and Engineering. A.V. is supported by a Simons Investigator grant and  by NSF-DMR 2220703. A.V. and R.V. are supported by the Simons Collaboration on Ultra-Quantum Matter, which is a grant from the Simons Foundation (618615, A.V.).
G.Z. is supported by the U.S. Department of Energy, Office of Science, National Quantum Information Science Research Centers, Co-design Center for Quantum Advantage (C2QA) under contract number DE-SC0012704.
We acknowledge the use of IBM Quantum services for this work.

\textbf{Author Contributions.} E.H.C. led the execution and analysis of the experimental data. G.Y.Z. and R.V. led the theoretical developments; G.Y.Z. developed all numerical simulations and additional experimental analysis code. A.S., E.B. and D.L. contributed technical insights and code related to characterizing the states. N.T., G.Z., A.V., and S.T. contributed theoretical insights related to the critical transition; S.T. provided additional insights related to entanglement generation and verification. S.S. and A.K. provided experimental support and access, and contributed to the design of the experiments. E.H.C., G.Y.Z., R.V. drafted the manuscript and supplemental material; all authors contributed to revising both.

\textbf{Data availability.} The data supporting the findings of this study will be available before publication.

\textbf{Code availability.} Simulation and data analysis code may be made available upon reasonable request.

\textbf{Competing interests.} The authors declare no competing interests.
\end{acknowledgements}

\section*{Methods}
\subsection*{127 superconducting qubit device}\label{method:device}
We performed all experiments on \textit{ibm\_sherbrooke}, a 127-qubit Eagle r3 processor. 
The entangling gate has a native $ZX$ interactions and is known as a Echoed Cross-Resonance (ECR) gate with a median error of $0.0077$, with a 50\% confidence interval of [0.006,~0.008]. The two-qubit gate times across the device were uniformly set to 533.3~nanoseconds,  similar to the method described in ~\cite{kim2023evidence}. 
The median square root of Pauli-X error rate was $0.0002$ [0.0002,~0.0004]. The readout error was $0.010$ [0.007,~0.021] with typical measurement times of $\approx1244.4$~nanoseconds. The qubits under study had a median $T1\approx293~\mu s$ and $T2\approx173~\mu s$. 
Circuits were executed on the device at a clock rate of 1kHz~\cite{sundaresan2020ecr}. For all data found in Figures 3-5, experimental error bars were calculated from the standard error on 20,000 shots at the 1kHz clock rate.

\begin{figure}[ht]
\includegraphics[width=\columnwidth]{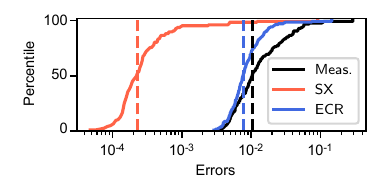}
\caption{
\textbf{
Typical error rates,} in cumulative format, for Echoed Cross Resonance (ECR, blue), square-root of Pauli-X (SX, red), and measurement (Meas, black) gates. Dashed lines represent medians of distributions.
}\label{fig:suppErrors}
\end{figure}

\subsection*{Decomposition of ZZ gates}\label{method:decomposition}
For fidelity comparisons in Fig.~\ref{fig:fig2}, we compiled the $ZZ(t=\pi/4)$ gate into a single ECR gate, which is the native basis gate on the device, and single-qubit rotations. For Fig.~\ref{fig:fig1}, the $ZZ(t)$ gates were all decomposed into two ECR gates with a virtual-$R_z(t)$ gate in between - resulting in a depth-6 unitary circuit followed by a layer of measurements. We note that further improvements could be accessed by shortening the $ZZ(t)$ gate time with fractional $ZX$ rotations~\cite{earnest2021pulse, kim2023scalable} that are accessible on the device. Hadamard gates, decomposed into two square root of Pauli-X gates and virtual-$R_z(t)$ gates, were also used for the preparation and readout of qubits.

\subsection*{Quantum Circuit transpilation}\label{method:transpilation}
For both unitary and measurement-based experiments, dynamical-decoupling (DD) was used in the same fashion. All single- and two-qubit gates were scheduled as late as possible after initialization in the ground state, and all idle periods after the first operations were replaced with a $X_{+\pi}-X_{-\pi}$ sequence in which the total idling period was divided proportionally according to a $1:2:1$ ratio surrounding the $X$ gates. Consequently, the unitary-based protocols benefited more from applying DD than for measurement-based circuits.

Furthermore, we identified at least 12 different ways to schedule entangling gates for the measurement-based, and found some schedules to significantly outperform others~\cite{supplement}. This is consistent with our expectation that certain gates, when executed in parallel, can induce frequency collisions on the device that reduce the fidelity of the entangling gates.

\subsection*{Monte Carlo Sampling \\of GHZ Stabilizer Observables}\label{method:stabilizers}
For size-$N$ GHZ states generated at the fixed point of the Nishimori line, we randomly measured up to $S$ different non-$Z$ stabilizers (e.g. weight-$N$ observables containing only $X$ and $Y$ Paulis). We also included measurements of the system qubits in the all-$Z$ basis, which allows us to reconstruct any of the $2^{N-1}$ possible $Z$-only stabilizers of the GHZ state. In practice, however, we perform a binomial re-sampling of $Z$-only and non-$Z$ stabilizers with equal probability to emulate the proposed Monte Carlo sampling method for fidelity state estimation~\cite{Flammia11fidelity,Silva11fidelity}. By sampling $2S'$ random instances from $2^N$
\begin{equation}
\begin{split}
\mathcal{F}  &= \frac{1}{2S'} \sum_{i=1}^{2S'} {\langle \mathcal{O}_i \rangle} \,.
\end{split}
\end{equation}
an average fidelity was estimated. The exact samples given $N$ sites, $S$ samples and $k$ resampling instances are, in $N$, $S$, $k$ notation: (10, 30, 500), (28, 75, 500), and (54, 100, 500). For each of the $k$ resampling instance, a total of $2S'$ observables were sampled but drawn randomly without replacement from the $S$ non-$Z$ stabilizers or from the 1000 $Z$-only stabilizers. Thus, $S'\approx 3S/4$ of the non-Z stabilizers were sampled without replacement from $S$, and another $S'\approx 3S/4$ $Z$-only stabilizers were sampled also without replacement from 1000 unique $Z$-only stabilizers. Both sets $S'$ combined to give $\approx 2S'$ observables as part of this binomial re-sampling procedure.

In all expectation values above, we randomly applied $X$-gates before readout of the system qubits and, after applying the correcting spin flip to sites on which $Z$ and $Y$ Paulis were supported, calculated the expectation values of the random stabilizers. Although readout was ``twirled'', the model-free readout correction was not applied~\cite{van2022model,supplement}.

\subsection*{Fits to noise model}

Two most basic linear observables are analytically known in the noiseless limit: $\langle W\rangle = \sin(2t_A)^6$, $\langle ZXZ\rangle = \sin(2 t_A)$. Let us consider two phenomenological errors: readout errors on the the auxiliary qubit measured in $X$ basis, with probability $p_s$; and that on the system qubit measured in $Z$ basis, with probability $p_\sigma$. Note that they also include the consequence of some of the mid-circuit bit-flip or phase-flip errors that propagate to yield the same effect in the end, such as the bit-flip (phase-flip) for system (auxiliary) qubits at the moment after the $R_{zz}$ gates. These two error rates turn the expectation values of the above observables into 
$\langle W\rangle = (1-2p_s)^6 \sin(2t_A)^6$, $\langle ZXZ\rangle =  (1-2p_s)(1-2p_\sigma)^2\sin(2 t_A)$. We can perform a linear fit to extract such phenomenological error rates per bond and plaquette, which are then averaged over the lattice for mean values and standard deviations. The averaged effective errors per qubit weakly grows with the total number of qubits in our three experimental implementations, as seen in Table.~\ref{table:pspsigma}.

\begin{table}[h]
\begin{tabular}{cccc}
\hline
 \begin{tabular}[c]{@{}c@{}}	System\\size\\$(N)$		\end{tabular} 
& \begin{tabular}[c]{@{}c@{}}	$L_y$				\end{tabular} 
& \begin{tabular}[c]{@{}c@{}}	$p_s$\\ (auxiliary)		\end{tabular} 
& \begin{tabular}[c]{@{}c@{}}	$p_{\sigma}$\\ (system)		\end{tabular} 
\\ 
\hline
10 & 2         & 0.042                                                       & 0.012                                                      \\ 
28 & 3         & 0.051                                                       & 0.018                                                      \\ 
54 & 4         & 0.056                                                       & 0.023                                                      \\ \hline
\end{tabular}
\caption{{\bf Two-parameter noise model.} Fits to experimental data gives $p_s$ which captures errors at the auxiliary qubits, and $p_\sigma$ at the system qubits.}
\label{table:pspsigma}
\end{table}

For the one-dimensional protocol where we do not have the Wilson loop $\langle W\rangle$, we can use two Wilson lines of different lengths, e.g. $\langle ZXZ\rangle $ and $\langle ZX\mathbb{I}XZ\rangle$ to extract the two parameters for auxiliary and system qubits, respectively.

\bibliography{measurements, 2022-nishi}

\begin{thebibliography}{52}%
\makeatletter
\providecommand \@ifxundefined [1]{%
 \@ifx{#1\undefined}
}%
\providecommand \@ifnum [1]{%
 \ifnum #1\expandafter \@firstoftwo
 \else \expandafter \@secondoftwo
 \fi
}%
\providecommand \@ifx [1]{%
 \ifx #1\expandafter \@firstoftwo
 \else \expandafter \@secondoftwo
 \fi
}%
\providecommand \natexlab [1]{#1}%
\providecommand \enquote  [1]{``#1''}%
\providecommand \bibnamefont  [1]{#1}%
\providecommand \bibfnamefont [1]{#1}%
\providecommand \citenamefont [1]{#1}%
\providecommand \href@noop [0]{\@secondoftwo}%
\providecommand \href [0]{\begingroup \@sanitize@url \@href}%
\providecommand \@href[1]{\@@startlink{#1}\@@href}%
\providecommand \@@href[1]{\endgroup#1\@@endlink}%
\providecommand \@sanitize@url [0]{\catcode `\\12\catcode `\$12\catcode
  `\&12\catcode `\#12\catcode `\^12\catcode `\_12\catcode `\%12\relax}%
\providecommand \@@startlink[1]{}%
\providecommand \@@endlink[0]{}%
\providecommand \url  [0]{\begingroup\@sanitize@url \@url }%
\providecommand \@url [1]{\endgroup\@href {#1}{\urlprefix }}%
\providecommand \urlprefix  [0]{URL }%
\providecommand \Eprint [0]{\href }%
\providecommand \doibase [0]{https://doi.org/}%
\providecommand \selectlanguage [0]{\@gobble}%
\providecommand \bibinfo  [0]{\@secondoftwo}%
\providecommand \bibfield  [0]{\@secondoftwo}%
\providecommand \translation [1]{[#1]}%
\providecommand \BibitemOpen [0]{}%
\providecommand \bibitemStop [0]{}%
\providecommand \bibitemNoStop [0]{.\EOS\space}%
\providecommand \EOS [0]{\spacefactor3000\relax}%
\providecommand \BibitemShut  [1]{\csname bibitem#1\endcsname}%
\let\auto@bib@innerbib\@empty
\bibitem [{\citenamefont {Bravyi}\ \emph {et~al.}(2006)\citenamefont {Bravyi},
  \citenamefont {Hastings},\ and\ \citenamefont
  {Verstraete}}]{BravyiHastingsVerstraete06}%
  \BibitemOpen
  \bibfield  {author} {\bibinfo {author} {\bibfnamefont {S.}~\bibnamefont
  {Bravyi}}, \bibinfo {author} {\bibfnamefont {M.~B.}\ \bibnamefont
  {Hastings}},\ and\ \bibinfo {author} {\bibfnamefont {F.}~\bibnamefont
  {Verstraete}},\ }\bibfield  {title} {\bibinfo {title} {{Lieb-Robinson Bounds
  and the Generation of Correlations and Topological Quantum Order}},\ }\href
  {https://doi.org/10.1103/PhysRevLett.97.050401} {\bibfield  {journal}
  {\bibinfo  {journal} {Phys. Rev. Lett.}\ }\textbf {\bibinfo {volume} {97}},\
  \bibinfo {pages} {050401} (\bibinfo {year} {2006})}\BibitemShut {NoStop}%
\bibitem [{\citenamefont {Hastings}(2012)}]{Hastings2010}%
  \BibitemOpen
  \bibfield  {author} {\bibinfo {author} {\bibfnamefont {M.~B.}\ \bibnamefont
  {Hastings}},\ }\bibfield  {title} {\bibinfo {title} {Locality in quantum
  systems},\ }in\ \href@noop {} {\emph {\bibinfo {booktitle} {Quantum Theory
  from Small to Large Scales}}},\ \bibinfo {series and number} {Les Houches
  2010, Session 95}\ (\bibinfo  {publisher} {Oxford University Press},\
  \bibinfo {year} {2012})\ pp.\ \bibinfo {pages} {171--212}\BibitemShut
  {NoStop}%
\bibitem [{\citenamefont {Chen}\ \emph
  {et~al.}(2011{\natexlab{a}})\citenamefont {Chen}, \citenamefont {Gu},\ and\
  \citenamefont {Wen}}]{ChenGuWen11A}%
  \BibitemOpen
  \bibfield  {author} {\bibinfo {author} {\bibfnamefont {X.}~\bibnamefont
  {Chen}}, \bibinfo {author} {\bibfnamefont {Z.-C.}\ \bibnamefont {Gu}},\ and\
  \bibinfo {author} {\bibfnamefont {X.-G.}\ \bibnamefont {Wen}},\ }\bibfield
  {title} {\bibinfo {title} {Classification of gapped symmetric phases in
  one-dimensional spin systems},\ }\href
  {https://doi.org/10.1103/PhysRevB.83.035107} {\bibfield  {journal} {\bibinfo
  {journal} {Phys. Rev. B}\ }\textbf {\bibinfo {volume} {83}},\ \bibinfo
  {pages} {035107} (\bibinfo {year} {2011}{\natexlab{a}})}\BibitemShut
  {NoStop}%
\bibitem [{\citenamefont {Chen}\ \emph
  {et~al.}(2011{\natexlab{b}})\citenamefont {Chen}, \citenamefont {Gu},\ and\
  \citenamefont {Wen}}]{ChenGuWen11B}%
  \BibitemOpen
  \bibfield  {author} {\bibinfo {author} {\bibfnamefont {X.}~\bibnamefont
  {Chen}}, \bibinfo {author} {\bibfnamefont {Z.-C.}\ \bibnamefont {Gu}},\ and\
  \bibinfo {author} {\bibfnamefont {X.-G.}\ \bibnamefont {Wen}},\ }\bibfield
  {title} {\bibinfo {title} {Complete classification of one-dimensional gapped
  quantum phases in interacting spin systems},\ }\href
  {https://doi.org/10.1103/PhysRevB.84.235128} {\bibfield  {journal} {\bibinfo
  {journal} {Phys. Rev. B}\ }\textbf {\bibinfo {volume} {84}},\ \bibinfo
  {pages} {235128} (\bibinfo {year} {2011}{\natexlab{b}})}\BibitemShut
  {NoStop}%
\bibitem [{\citenamefont {Zeng}\ and\ \citenamefont {Wen}(2015)}]{ZengWen15}%
  \BibitemOpen
  \bibfield  {author} {\bibinfo {author} {\bibfnamefont {B.}~\bibnamefont
  {Zeng}}\ and\ \bibinfo {author} {\bibfnamefont {X.-G.}\ \bibnamefont {Wen}},\
  }\bibfield  {title} {\bibinfo {title} {Gapped quantum liquids and topological
  order, stochastic local transformations and emergence of unitarity},\ }\href
  {https://doi.org/10.1103/PhysRevB.91.125121} {\bibfield  {journal} {\bibinfo
  {journal} {Phys. Rev. B}\ }\textbf {\bibinfo {volume} {91}},\ \bibinfo
  {pages} {125121} (\bibinfo {year} {2015})}\BibitemShut {NoStop}%
\bibitem [{\citenamefont {Huang}\ and\ \citenamefont
  {Chen}(2015)}]{HuangChen2015}%
  \BibitemOpen
  \bibfield  {author} {\bibinfo {author} {\bibfnamefont {Y.}~\bibnamefont
  {Huang}}\ and\ \bibinfo {author} {\bibfnamefont {X.}~\bibnamefont {Chen}},\
  }\bibfield  {title} {\bibinfo {title} {Quantum circuit complexity of
  one-dimensional topological phases},\ }\href
  {https://doi.org/10.1103/PhysRevB.91.195143} {\bibfield  {journal} {\bibinfo
  {journal} {Phys. Rev. B}\ }\textbf {\bibinfo {volume} {91}},\ \bibinfo
  {pages} {195143} (\bibinfo {year} {2015})}\BibitemShut {NoStop}%
\bibitem [{\citenamefont {Haah}(2016)}]{Haah2016}%
  \BibitemOpen
  \bibfield  {author} {\bibinfo {author} {\bibfnamefont {J.}~\bibnamefont
  {Haah}},\ }\bibfield  {title} {\bibinfo {title} {An invariant of
  topologically ordered states under local unitary transformations},\ }\href
  {https://doi.org/10.1007/s00220-016-2594-y} {\bibfield  {journal} {\bibinfo
  {journal} {Communications in Mathematical Physics}\ }\textbf {\bibinfo
  {volume} {342}},\ \bibinfo {pages} {771} (\bibinfo {year}
  {2016})}\BibitemShut {NoStop}%
\bibitem [{\citenamefont {Mooney}\ \emph
  {et~al.}(2021{\natexlab{a}})\citenamefont {Mooney}, \citenamefont {White},
  \citenamefont {Hill},\ and\ \citenamefont {Hollenberg}}]{mooney2021whole}%
  \BibitemOpen
  \bibfield  {author} {\bibinfo {author} {\bibfnamefont {G.~J.}\ \bibnamefont
  {Mooney}}, \bibinfo {author} {\bibfnamefont {G.~A.~L.}\ \bibnamefont
  {White}}, \bibinfo {author} {\bibfnamefont {C.~D.}\ \bibnamefont {Hill}},\
  and\ \bibinfo {author} {\bibfnamefont {L.~C.~L.}\ \bibnamefont
  {Hollenberg}},\ }\bibfield  {title} {\bibinfo {title} {{Whole-Device
  Entanglement in a 65-Qubit Superconducting Quantum Computer}},\ }\href
  {https://doi.org/10.1002/qute.202100061} {\bibfield  {journal} {\bibinfo
  {journal} {Advanced Quantum Technologies}\ }\textbf {\bibinfo {volume} {4}},\
  \bibinfo {pages} {2100061} (\bibinfo {year}
  {2021}{\natexlab{a}})}\BibitemShut {NoStop}%
\bibitem [{\citenamefont {Mooney}\ \emph
  {et~al.}(2021{\natexlab{b}})\citenamefont {Mooney}, \citenamefont {White},
  \citenamefont {Hill},\ and\ \citenamefont
  {Hollenberg}}]{mooney_generation_2021}%
  \BibitemOpen
  \bibfield  {author} {\bibinfo {author} {\bibfnamefont {G.~J.}\ \bibnamefont
  {Mooney}}, \bibinfo {author} {\bibfnamefont {G.~A.~L.}\ \bibnamefont
  {White}}, \bibinfo {author} {\bibfnamefont {C.~D.}\ \bibnamefont {Hill}},\
  and\ \bibinfo {author} {\bibfnamefont {L.~C.~L.}\ \bibnamefont
  {Hollenberg}},\ }\bibfield  {title} {\bibinfo {title} {Generation and
  verification of 27-qubit {Greenberger}-{Horne}-{Zeilinger} states in a
  superconducting quantum computer},\ }\href
  {https://doi.org/10.1088/2399-6528/ac1df7} {\bibfield  {journal} {\bibinfo
  {journal} {Journal of Physics Communications}\ }\textbf {\bibinfo {volume}
  {5}},\ \bibinfo {pages} {095004} (\bibinfo {year}
  {2021}{\natexlab{b}})}\BibitemShut {NoStop}%
\bibitem [{\citenamefont {Wei}\ \emph {et~al.}(2020)\citenamefont {Wei},
  \citenamefont {Lauer}, \citenamefont {Srinivasan}, \citenamefont
  {Sundaresan}, \citenamefont {McClure}, \citenamefont {Toyli}, \citenamefont
  {McKay}, \citenamefont {Gambetta},\ and\ \citenamefont
  {Sheldon}}]{wei2020verifying}%
  \BibitemOpen
  \bibfield  {author} {\bibinfo {author} {\bibfnamefont {K.~X.}\ \bibnamefont
  {Wei}}, \bibinfo {author} {\bibfnamefont {I.}~\bibnamefont {Lauer}}, \bibinfo
  {author} {\bibfnamefont {S.}~\bibnamefont {Srinivasan}}, \bibinfo {author}
  {\bibfnamefont {N.}~\bibnamefont {Sundaresan}}, \bibinfo {author}
  {\bibfnamefont {D.~T.}\ \bibnamefont {McClure}}, \bibinfo {author}
  {\bibfnamefont {D.}~\bibnamefont {Toyli}}, \bibinfo {author} {\bibfnamefont
  {D.~C.}\ \bibnamefont {McKay}}, \bibinfo {author} {\bibfnamefont {J.~M.}\
  \bibnamefont {Gambetta}},\ and\ \bibinfo {author} {\bibfnamefont
  {S.}~\bibnamefont {Sheldon}},\ }\bibfield  {title} {\bibinfo {title}
  {{Verifying multipartite entangled Greenberger-Horne-Zeilinger states via
  multiple quantum coherences}},\ }\href
  {https://doi.org/10.1103/PhysRevA.101.032343} {\bibfield  {journal} {\bibinfo
   {journal} {Phys. Rev. A}\ }\textbf {\bibinfo {volume} {101}},\ \bibinfo
  {pages} {032343} (\bibinfo {year} {2020})}\BibitemShut {NoStop}%
\bibitem [{\citenamefont {Briegel}\ and\ \citenamefont
  {Raussendorf}(2001)}]{Raussendorf2001}%
  \BibitemOpen
  \bibfield  {author} {\bibinfo {author} {\bibfnamefont {H.~J.}\ \bibnamefont
  {Briegel}}\ and\ \bibinfo {author} {\bibfnamefont {R.}~\bibnamefont
  {Raussendorf}},\ }\bibfield  {title} {\bibinfo {title} {{Persistent
  Entanglement in Arrays of Interacting Particles}},\ }\href
  {https://doi.org/10.1103/PhysRevLett.86.910} {\bibfield  {journal} {\bibinfo
  {journal} {Phys. Rev. Lett.}\ }\textbf {\bibinfo {volume} {86}},\ \bibinfo
  {pages} {910} (\bibinfo {year} {2001})}\BibitemShut {NoStop}%
\bibitem [{\citenamefont {Raussendorf}\ \emph {et~al.}(2005)\citenamefont
  {Raussendorf}, \citenamefont {Bravyi},\ and\ \citenamefont
  {Harrington}}]{Raussendorf2005}%
  \BibitemOpen
  \bibfield  {author} {\bibinfo {author} {\bibfnamefont {R.}~\bibnamefont
  {Raussendorf}}, \bibinfo {author} {\bibfnamefont {S.}~\bibnamefont
  {Bravyi}},\ and\ \bibinfo {author} {\bibfnamefont {J.}~\bibnamefont
  {Harrington}},\ }\bibfield  {title} {\bibinfo {title} {Long-range quantum
  entanglement in noisy cluster states},\ }\href
  {https://doi.org/10.1103/PhysRevA.71.062313} {\bibfield  {journal} {\bibinfo
  {journal} {Phys. Rev. A}\ }\textbf {\bibinfo {volume} {71}},\ \bibinfo
  {pages} {062313} (\bibinfo {year} {2005})}\BibitemShut {NoStop}%
\bibitem [{\citenamefont {Piroli}\ \emph {et~al.}(2021)\citenamefont {Piroli},
  \citenamefont {Styliaris},\ and\ \citenamefont
  {Cirac}}]{PhysRevLett.127.220503}%
  \BibitemOpen
  \bibfield  {author} {\bibinfo {author} {\bibfnamefont {L.}~\bibnamefont
  {Piroli}}, \bibinfo {author} {\bibfnamefont {G.}~\bibnamefont {Styliaris}},\
  and\ \bibinfo {author} {\bibfnamefont {J.~I.}\ \bibnamefont {Cirac}},\
  }\bibfield  {title} {\bibinfo {title} {Quantum circuits assisted by local
  operations and classical communication: Transformations and phases of
  matter},\ }\href {https://doi.org/10.1103/PhysRevLett.127.220503} {\bibfield
  {journal} {\bibinfo  {journal} {Phys. Rev. Lett.}\ }\textbf {\bibinfo
  {volume} {127}},\ \bibinfo {pages} {220503} (\bibinfo {year}
  {2021})}\BibitemShut {NoStop}%
\bibitem [{\citenamefont {Tantivasadakarn}\ \emph {et~al.}(2021)\citenamefont
  {Tantivasadakarn}, \citenamefont {Thorngren}, \citenamefont {Vishwanath},\
  and\ \citenamefont {Verresen}}]{Nat2021measure}%
  \BibitemOpen
  \bibfield  {author} {\bibinfo {author} {\bibfnamefont {N.}~\bibnamefont
  {Tantivasadakarn}}, \bibinfo {author} {\bibfnamefont {R.}~\bibnamefont
  {Thorngren}}, \bibinfo {author} {\bibfnamefont {A.}~\bibnamefont
  {Vishwanath}},\ and\ \bibinfo {author} {\bibfnamefont {R.}~\bibnamefont
  {Verresen}},\ }\bibfield  {title} {\bibinfo {title} {Long-range entanglement
  from measuring symmetry-protected topological phases},\ }\href@noop {} {\
  (\bibinfo {year} {2021})},\ \Eprint {https://arxiv.org/abs/2112.01519}
  {arXiv:2112.01519} \BibitemShut {NoStop}%
\bibitem [{\citenamefont {Verresen}\ \emph {et~al.}(2021)\citenamefont
  {Verresen}, \citenamefont {Tantivasadakarn},\ and\ \citenamefont
  {Vishwanath}}]{Verresen2021cat}%
  \BibitemOpen
  \bibfield  {author} {\bibinfo {author} {\bibfnamefont {R.}~\bibnamefont
  {Verresen}}, \bibinfo {author} {\bibfnamefont {N.}~\bibnamefont
  {Tantivasadakarn}},\ and\ \bibinfo {author} {\bibfnamefont {A.}~\bibnamefont
  {Vishwanath}},\ }\bibfield  {title} {\bibinfo {title} {{Efficiently preparing
  Schr{\"o}dinger’s cat, fractons and non-Abelian topological order in
  quantum devices}},\ }\href@noop {} {\  (\bibinfo {year} {2021})},\ \Eprint
  {https://arxiv.org/abs/2112.03061} {arXiv:2112.03061} \BibitemShut {NoStop}%
\bibitem [{\citenamefont {Bravyi}\ \emph {et~al.}(2022)\citenamefont {Bravyi},
  \citenamefont {Kim}, \citenamefont {Kliesch},\ and\ \citenamefont
  {Koenig}}]{Bravyi2022}%
  \BibitemOpen
  \bibfield  {author} {\bibinfo {author} {\bibfnamefont {S.}~\bibnamefont
  {Bravyi}}, \bibinfo {author} {\bibfnamefont {I.}~\bibnamefont {Kim}},
  \bibinfo {author} {\bibfnamefont {A.}~\bibnamefont {Kliesch}},\ and\ \bibinfo
  {author} {\bibfnamefont {R.}~\bibnamefont {Koenig}},\ }\bibfield  {title}
  {\bibinfo {title} {Adaptive constant-depth circuits for manipulating
  non-abelian anyons},\ }\href@noop {} {\  (\bibinfo {year} {2022})},\ \Eprint
  {https://arxiv.org/abs/2205.01933} {arXiv:2205.01933} \BibitemShut {NoStop}%
\bibitem [{\citenamefont {Lu}\ \emph {et~al.}(2022)\citenamefont {Lu},
  \citenamefont {Lessa}, \citenamefont {Kim},\ and\ \citenamefont
  {Hsieh}}]{Lu22}%
  \BibitemOpen
  \bibfield  {author} {\bibinfo {author} {\bibfnamefont {T.-C.}\ \bibnamefont
  {Lu}}, \bibinfo {author} {\bibfnamefont {L.~A.}\ \bibnamefont {Lessa}},
  \bibinfo {author} {\bibfnamefont {I.~H.}\ \bibnamefont {Kim}},\ and\ \bibinfo
  {author} {\bibfnamefont {T.~H.}\ \bibnamefont {Hsieh}},\ }\bibfield  {title}
  {\bibinfo {title} {{Measurement as a Shortcut to Long-Range Entangled Quantum
  Matter}},\ }\href {https://doi.org/10.1103/PRXQuantum.3.040337} {\bibfield
  {journal} {\bibinfo  {journal} {PRX Quantum}\ }\textbf {\bibinfo {volume}
  {3}},\ \bibinfo {pages} {040337} (\bibinfo {year} {2022})}\BibitemShut
  {NoStop}%
\bibitem [{\citenamefont {Zhu}\ \emph {et~al.}(2022)\citenamefont {Zhu},
  \citenamefont {Tantivasadakarn}, \citenamefont {Vishwanath}, \citenamefont
  {Trebst},\ and\ \citenamefont {Verresen}}]{zhu_nishimoris_2022}%
  \BibitemOpen
  \bibfield  {author} {\bibinfo {author} {\bibfnamefont {G.-Y.}\ \bibnamefont
  {Zhu}}, \bibinfo {author} {\bibfnamefont {N.}~\bibnamefont
  {Tantivasadakarn}}, \bibinfo {author} {\bibfnamefont {A.}~\bibnamefont
  {Vishwanath}}, \bibinfo {author} {\bibfnamefont {S.}~\bibnamefont {Trebst}},\
  and\ \bibinfo {author} {\bibfnamefont {R.}~\bibnamefont {Verresen}},\
  }\bibfield  {title} {\bibinfo {title} {{Nishimori's cat: stable long-range
  entanglement from finite-depth unitaries and weak measurements}},\
  }\href@noop {} {\  (\bibinfo {year} {2022})},\ \Eprint
  {https://arxiv.org/abs/2208.11136} {arXiv:2208.11136} \BibitemShut {NoStop}%
\bibitem [{\citenamefont {Tantivasadakarn}\ \emph
  {et~al.}(2023{\natexlab{a}})\citenamefont {Tantivasadakarn}, \citenamefont
  {Vishwanath},\ and\ \citenamefont
  {Verresen}}]{tantivasadakarn_hierarchy_2023}%
  \BibitemOpen
  \bibfield  {author} {\bibinfo {author} {\bibfnamefont {N.}~\bibnamefont
  {Tantivasadakarn}}, \bibinfo {author} {\bibfnamefont {A.}~\bibnamefont
  {Vishwanath}},\ and\ \bibinfo {author} {\bibfnamefont {R.}~\bibnamefont
  {Verresen}},\ }\bibfield  {title} {\bibinfo {title} {Hierarchy of
  {Topological} {Order} {From} {Finite}-{Depth} {Unitaries}, {Measurement}, and
  {Feedforward}},\ }\href {https://doi.org/10.1103/PRXQuantum.4.020339}
  {\bibfield  {journal} {\bibinfo  {journal} {PRX Quantum}\ }\textbf {\bibinfo
  {volume} {4}},\ \bibinfo {pages} {020339} (\bibinfo {year}
  {2023}{\natexlab{a}})}\BibitemShut {NoStop}%
\bibitem [{\citenamefont {Tantivasadakarn}\ \emph
  {et~al.}(2023{\natexlab{b}})\citenamefont {Tantivasadakarn}, \citenamefont
  {Verresen},\ and\ \citenamefont {Vishwanath}}]{PhysRevLett.131.060405}%
  \BibitemOpen
  \bibfield  {author} {\bibinfo {author} {\bibfnamefont {N.}~\bibnamefont
  {Tantivasadakarn}}, \bibinfo {author} {\bibfnamefont {R.}~\bibnamefont
  {Verresen}},\ and\ \bibinfo {author} {\bibfnamefont {A.}~\bibnamefont
  {Vishwanath}},\ }\bibfield  {title} {\bibinfo {title} {Shortest route to
  non-abelian topological order on a quantum processor},\ }\href
  {https://doi.org/10.1103/PhysRevLett.131.060405} {\bibfield  {journal}
  {\bibinfo  {journal} {Phys. Rev. Lett.}\ }\textbf {\bibinfo {volume} {131}},\
  \bibinfo {pages} {060405} (\bibinfo {year} {2023}{\natexlab{b}})}\BibitemShut
  {NoStop}%
\bibitem [{\citenamefont {{Lee}}\ \emph {et~al.}(2022)\citenamefont {{Lee}},
  \citenamefont {{Ji}}, \citenamefont {{Bi}},\ and\ \citenamefont
  {{Fisher}}}]{Lee22}%
  \BibitemOpen
  \bibfield  {author} {\bibinfo {author} {\bibfnamefont {J.~Y.}\ \bibnamefont
  {{Lee}}}, \bibinfo {author} {\bibfnamefont {W.}~\bibnamefont {{Ji}}},
  \bibinfo {author} {\bibfnamefont {Z.}~\bibnamefont {{Bi}}},\ and\ \bibinfo
  {author} {\bibfnamefont {M.~P.~A.}\ \bibnamefont {{Fisher}}},\ }\bibfield
  {title} {\bibinfo {title} {{Decoding Measurement-Prepared Quantum Phases and
  Transitions: from Ising model to gauge theory, and beyond}},\ }\href@noop {}
  {\  (\bibinfo {year} {2022})},\ \Eprint {https://arxiv.org/abs/2208.11699}
  {arXiv:2208.11699} \BibitemShut {NoStop}%
\bibitem [{\citenamefont {Li}\ \emph {et~al.}(2023)\citenamefont {Li},
  \citenamefont {Sukeno}, \citenamefont {Mana}, \citenamefont {Nautrup},\ and\
  \citenamefont {Wei}}]{li2023symmetryenriched}%
  \BibitemOpen
  \bibfield  {author} {\bibinfo {author} {\bibfnamefont {Y.}~\bibnamefont
  {Li}}, \bibinfo {author} {\bibfnamefont {H.}~\bibnamefont {Sukeno}}, \bibinfo
  {author} {\bibfnamefont {A.~P.}\ \bibnamefont {Mana}}, \bibinfo {author}
  {\bibfnamefont {H.~P.}\ \bibnamefont {Nautrup}},\ and\ \bibinfo {author}
  {\bibfnamefont {T.-C.}\ \bibnamefont {Wei}},\ }\href@noop {} {\bibinfo
  {title} {Symmetry-enriched topological order from partially gauging
  symmetry-protected topologically ordered states assisted by measurements}}
  (\bibinfo {year} {2023}),\ \Eprint {https://arxiv.org/abs/2305.09747}
  {arXiv:2305.09747 [quant-ph]} \BibitemShut {NoStop}%
\bibitem [{\citenamefont {Buhrman}\ \emph {et~al.}(2023)\citenamefont
  {Buhrman}, \citenamefont {Folkertsma}, \citenamefont {Loff},\ and\
  \citenamefont {Neumann}}]{buhrman2023state}%
  \BibitemOpen
  \bibfield  {author} {\bibinfo {author} {\bibfnamefont {H.}~\bibnamefont
  {Buhrman}}, \bibinfo {author} {\bibfnamefont {M.}~\bibnamefont {Folkertsma}},
  \bibinfo {author} {\bibfnamefont {B.}~\bibnamefont {Loff}},\ and\ \bibinfo
  {author} {\bibfnamefont {N.~M.~P.}\ \bibnamefont {Neumann}},\ }\href@noop {}
  {\bibinfo {title} {State preparation by shallow circuits using feed forward}}
  (\bibinfo {year} {2023}),\ \Eprint {https://arxiv.org/abs/2307.14840}
  {arXiv:2307.14840 [quant-ph]} \BibitemShut {NoStop}%
\bibitem [{\citenamefont {Lu}\ \emph {et~al.}(2023)\citenamefont {Lu},
  \citenamefont {Zhang}, \citenamefont {Vijay},\ and\ \citenamefont
  {Hsieh}}]{lu_mixed-state_2023}%
  \BibitemOpen
  \bibfield  {author} {\bibinfo {author} {\bibfnamefont {T.-C.}\ \bibnamefont
  {Lu}}, \bibinfo {author} {\bibfnamefont {Z.}~\bibnamefont {Zhang}}, \bibinfo
  {author} {\bibfnamefont {S.}~\bibnamefont {Vijay}},\ and\ \bibinfo {author}
  {\bibfnamefont {T.~H.}\ \bibnamefont {Hsieh}},\ }\href@noop {} {\bibinfo
  {title} {Mixed-state long-range order and criticality from measurement and
  feedback}} (\bibinfo {year} {2023}),\ \Eprint
  {https://arxiv.org/abs/2303.15507} {arXiv:2303.15507} \BibitemShut {NoStop}%
\bibitem [{\citenamefont {Friedman}\ \emph {et~al.}(2023)\citenamefont
  {Friedman}, \citenamefont {Yin}, \citenamefont {Hong},\ and\ \citenamefont
  {Lucas}}]{friedman2023locality}%
  \BibitemOpen
  \bibfield  {author} {\bibinfo {author} {\bibfnamefont {A.~J.}\ \bibnamefont
  {Friedman}}, \bibinfo {author} {\bibfnamefont {C.}~\bibnamefont {Yin}},
  \bibinfo {author} {\bibfnamefont {Y.}~\bibnamefont {Hong}},\ and\ \bibinfo
  {author} {\bibfnamefont {A.}~\bibnamefont {Lucas}},\ }\bibfield  {title}
  {\bibinfo {title} {Locality and error correction in quantum dynamics with
  measurement},\ }\href@noop {} {\  (\bibinfo {year} {2023})},\ \Eprint
  {https://arxiv.org/abs/2206.09929} {arXiv:2206.09929} \BibitemShut {NoStop}%
\bibitem [{\citenamefont {Moses}\ \emph {et~al.}(2023)\citenamefont {Moses},
  \citenamefont {Baldwin}, \citenamefont {Allman}, \citenamefont {Ancona},
  \citenamefont {Ascarrunz}, \citenamefont {Barnes}, \citenamefont
  {Bartolotta}, \citenamefont {Bjork}, \citenamefont {Blanchard}, \citenamefont
  {Bohn}, \citenamefont {Bohnet}, \citenamefont {Brown}, \citenamefont
  {Burdick}, \citenamefont {Burton}, \citenamefont {Campbell}, \citenamefont
  {Campora~III}, \citenamefont {Carron}, \citenamefont {Chambers},
  \citenamefont {Chen}, \citenamefont {Chen}, \citenamefont {Chernoguzov},
  \citenamefont {Chertkov}, \citenamefont {Colina}, \citenamefont {DeCross},
  \citenamefont {Dreiling}, \citenamefont {Ertsgaard}, \citenamefont
  {Esposito}, \citenamefont {Estey}, \citenamefont {Fabrikant}, \citenamefont
  {Figgatt}, \citenamefont {Foltz}, \citenamefont {Foss-Feig}, \citenamefont
  {Francois}, \citenamefont {Gaebler}, \citenamefont {Gatterman}, \citenamefont
  {Gilbreth}, \citenamefont {Giles}, \citenamefont {Glynn}, \citenamefont
  {Hall}, \citenamefont {Hankin}, \citenamefont {Hansen}, \citenamefont
  {Hayes}, \citenamefont {Higashi}, \citenamefont {Hoffman}, \citenamefont
  {Horning}, \citenamefont {Hout}, \citenamefont {Jacobs}, \citenamefont
  {Johansen}, \citenamefont {Klein}, \citenamefont {Lauria}, \citenamefont
  {Lee}, \citenamefont {Liefer}, \citenamefont {Lu}, \citenamefont {Lucchetti},
  \citenamefont {Malm}, \citenamefont {Matheny}, \citenamefont {Mathewson},
  \citenamefont {Mayer}, \citenamefont {Miller}, \citenamefont {Mills},
  \citenamefont {Neyenhuis}, \citenamefont {Nugent}, \citenamefont {Olson},
  \citenamefont {Parks}, \citenamefont {Price}, \citenamefont {Price},
  \citenamefont {Pugh}, \citenamefont {Ransford}, \citenamefont {Reed},
  \citenamefont {Roman}, \citenamefont {Rowe}, \citenamefont {Ryan-Anderson},
  \citenamefont {Sanders}, \citenamefont {Sedlacek}, \citenamefont {Shevchuk},
  \citenamefont {Siegfried}, \citenamefont {Skripka}, \citenamefont {Spaun},
  \citenamefont {Sprenkle}, \citenamefont {Stutz}, \citenamefont {Swallows},
  \citenamefont {Tobey}, \citenamefont {Tran}, \citenamefont {Tran},
  \citenamefont {Vogt}, \citenamefont {Volin}, \citenamefont {Walker},
  \citenamefont {Zolot},\ and\ \citenamefont {Pino}}]{moses_race_2023}%
  \BibitemOpen
  \bibfield  {author} {\bibinfo {author} {\bibfnamefont {S.~A.}\ \bibnamefont
  {Moses}}, \bibinfo {author} {\bibfnamefont {C.~H.}\ \bibnamefont {Baldwin}},
  \bibinfo {author} {\bibfnamefont {M.~S.}\ \bibnamefont {Allman}}, \bibinfo
  {author} {\bibfnamefont {R.}~\bibnamefont {Ancona}}, \bibinfo {author}
  {\bibfnamefont {L.}~\bibnamefont {Ascarrunz}}, \bibinfo {author}
  {\bibfnamefont {C.}~\bibnamefont {Barnes}}, \bibinfo {author} {\bibfnamefont
  {J.}~\bibnamefont {Bartolotta}}, \bibinfo {author} {\bibfnamefont
  {B.}~\bibnamefont {Bjork}}, \bibinfo {author} {\bibfnamefont
  {P.}~\bibnamefont {Blanchard}}, \bibinfo {author} {\bibfnamefont
  {M.}~\bibnamefont {Bohn}}, \bibinfo {author} {\bibfnamefont {J.~G.}\
  \bibnamefont {Bohnet}}, \bibinfo {author} {\bibfnamefont {N.~C.}\
  \bibnamefont {Brown}}, \bibinfo {author} {\bibfnamefont {N.~Q.}\ \bibnamefont
  {Burdick}}, \bibinfo {author} {\bibfnamefont {W.~C.}\ \bibnamefont {Burton}},
  \bibinfo {author} {\bibfnamefont {S.~L.}\ \bibnamefont {Campbell}}, \bibinfo
  {author} {\bibfnamefont {J.~P.}\ \bibnamefont {Campora~III}}, \bibinfo
  {author} {\bibfnamefont {C.}~\bibnamefont {Carron}}, \bibinfo {author}
  {\bibfnamefont {J.}~\bibnamefont {Chambers}}, \bibinfo {author}
  {\bibfnamefont {J.~W.}\ \bibnamefont {Chen}}, \bibinfo {author}
  {\bibfnamefont {Y.~H.}\ \bibnamefont {Chen}}, \bibinfo {author}
  {\bibfnamefont {A.}~\bibnamefont {Chernoguzov}}, \bibinfo {author}
  {\bibfnamefont {E.}~\bibnamefont {Chertkov}}, \bibinfo {author}
  {\bibfnamefont {J.}~\bibnamefont {Colina}}, \bibinfo {author} {\bibfnamefont
  {M.}~\bibnamefont {DeCross}}, \bibinfo {author} {\bibfnamefont {J.~M.}\
  \bibnamefont {Dreiling}}, \bibinfo {author} {\bibfnamefont {C.~T.}\
  \bibnamefont {Ertsgaard}}, \bibinfo {author} {\bibfnamefont {J.}~\bibnamefont
  {Esposito}}, \bibinfo {author} {\bibfnamefont {B.}~\bibnamefont {Estey}},
  \bibinfo {author} {\bibfnamefont {M.}~\bibnamefont {Fabrikant}}, \bibinfo
  {author} {\bibfnamefont {C.}~\bibnamefont {Figgatt}}, \bibinfo {author}
  {\bibfnamefont {C.}~\bibnamefont {Foltz}}, \bibinfo {author} {\bibfnamefont
  {M.}~\bibnamefont {Foss-Feig}}, \bibinfo {author} {\bibfnamefont
  {D.}~\bibnamefont {Francois}}, \bibinfo {author} {\bibfnamefont {J.~P.}\
  \bibnamefont {Gaebler}}, \bibinfo {author} {\bibfnamefont {T.~M.}\
  \bibnamefont {Gatterman}}, \bibinfo {author} {\bibfnamefont {C.~N.}\
  \bibnamefont {Gilbreth}}, \bibinfo {author} {\bibfnamefont {J.}~\bibnamefont
  {Giles}}, \bibinfo {author} {\bibfnamefont {E.}~\bibnamefont {Glynn}},
  \bibinfo {author} {\bibfnamefont {A.}~\bibnamefont {Hall}}, \bibinfo {author}
  {\bibfnamefont {A.~M.}\ \bibnamefont {Hankin}}, \bibinfo {author}
  {\bibfnamefont {A.}~\bibnamefont {Hansen}}, \bibinfo {author} {\bibfnamefont
  {D.}~\bibnamefont {Hayes}}, \bibinfo {author} {\bibfnamefont
  {B.}~\bibnamefont {Higashi}}, \bibinfo {author} {\bibfnamefont {I.~M.}\
  \bibnamefont {Hoffman}}, \bibinfo {author} {\bibfnamefont {B.}~\bibnamefont
  {Horning}}, \bibinfo {author} {\bibfnamefont {J.~J.}\ \bibnamefont {Hout}},
  \bibinfo {author} {\bibfnamefont {R.}~\bibnamefont {Jacobs}}, \bibinfo
  {author} {\bibfnamefont {J.}~\bibnamefont {Johansen}}, \bibinfo {author}
  {\bibfnamefont {T.}~\bibnamefont {Klein}}, \bibinfo {author} {\bibfnamefont
  {P.}~\bibnamefont {Lauria}}, \bibinfo {author} {\bibfnamefont
  {P.}~\bibnamefont {Lee}}, \bibinfo {author} {\bibfnamefont {D.}~\bibnamefont
  {Liefer}}, \bibinfo {author} {\bibfnamefont {S.~T.}\ \bibnamefont {Lu}},
  \bibinfo {author} {\bibfnamefont {D.}~\bibnamefont {Lucchetti}}, \bibinfo
  {author} {\bibfnamefont {A.}~\bibnamefont {Malm}}, \bibinfo {author}
  {\bibfnamefont {M.}~\bibnamefont {Matheny}}, \bibinfo {author} {\bibfnamefont
  {B.}~\bibnamefont {Mathewson}}, \bibinfo {author} {\bibfnamefont
  {K.}~\bibnamefont {Mayer}}, \bibinfo {author} {\bibfnamefont {D.~B.}\
  \bibnamefont {Miller}}, \bibinfo {author} {\bibfnamefont {M.}~\bibnamefont
  {Mills}}, \bibinfo {author} {\bibfnamefont {B.}~\bibnamefont {Neyenhuis}},
  \bibinfo {author} {\bibfnamefont {L.}~\bibnamefont {Nugent}}, \bibinfo
  {author} {\bibfnamefont {S.}~\bibnamefont {Olson}}, \bibinfo {author}
  {\bibfnamefont {J.}~\bibnamefont {Parks}}, \bibinfo {author} {\bibfnamefont
  {G.~N.}\ \bibnamefont {Price}}, \bibinfo {author} {\bibfnamefont
  {Z.}~\bibnamefont {Price}}, \bibinfo {author} {\bibfnamefont
  {M.}~\bibnamefont {Pugh}}, \bibinfo {author} {\bibfnamefont {A.}~\bibnamefont
  {Ransford}}, \bibinfo {author} {\bibfnamefont {A.~P.}\ \bibnamefont {Reed}},
  \bibinfo {author} {\bibfnamefont {C.}~\bibnamefont {Roman}}, \bibinfo
  {author} {\bibfnamefont {M.}~\bibnamefont {Rowe}}, \bibinfo {author}
  {\bibfnamefont {C.}~\bibnamefont {Ryan-Anderson}}, \bibinfo {author}
  {\bibfnamefont {S.}~\bibnamefont {Sanders}}, \bibinfo {author} {\bibfnamefont
  {J.}~\bibnamefont {Sedlacek}}, \bibinfo {author} {\bibfnamefont
  {P.}~\bibnamefont {Shevchuk}}, \bibinfo {author} {\bibfnamefont
  {P.}~\bibnamefont {Siegfried}}, \bibinfo {author} {\bibfnamefont
  {T.}~\bibnamefont {Skripka}}, \bibinfo {author} {\bibfnamefont
  {B.}~\bibnamefont {Spaun}}, \bibinfo {author} {\bibfnamefont {R.~T.}\
  \bibnamefont {Sprenkle}}, \bibinfo {author} {\bibfnamefont {R.~P.}\
  \bibnamefont {Stutz}}, \bibinfo {author} {\bibfnamefont {M.}~\bibnamefont
  {Swallows}}, \bibinfo {author} {\bibfnamefont {R.~I.}\ \bibnamefont {Tobey}},
  \bibinfo {author} {\bibfnamefont {A.}~\bibnamefont {Tran}}, \bibinfo {author}
  {\bibfnamefont {T.}~\bibnamefont {Tran}}, \bibinfo {author} {\bibfnamefont
  {E.}~\bibnamefont {Vogt}}, \bibinfo {author} {\bibfnamefont {C.}~\bibnamefont
  {Volin}}, \bibinfo {author} {\bibfnamefont {J.}~\bibnamefont {Walker}},
  \bibinfo {author} {\bibfnamefont {A.~M.}\ \bibnamefont {Zolot}},\ and\
  \bibinfo {author} {\bibfnamefont {J.~M.}\ \bibnamefont {Pino}},\ }\bibfield
  {title} {\bibinfo {title} {A {Race} {Track} {Trapped}-{Ion} {Quantum}
  {Processor}},\ }\href@noop {} {\  (\bibinfo {year} {2023})},\ \Eprint
  {https://arxiv.org/abs/2305.03828} {arXiv:2305.03828} \BibitemShut {NoStop}%
\bibitem [{\citenamefont {Dennis}\ \emph {et~al.}(2002)\citenamefont {Dennis},
  \citenamefont {Kitaev}, \citenamefont {Landahl},\ and\ \citenamefont
  {Preskill}}]{Preskill2002}%
  \BibitemOpen
  \bibfield  {author} {\bibinfo {author} {\bibfnamefont {E.}~\bibnamefont
  {Dennis}}, \bibinfo {author} {\bibfnamefont {A.}~\bibnamefont {Kitaev}},
  \bibinfo {author} {\bibfnamefont {A.}~\bibnamefont {Landahl}},\ and\ \bibinfo
  {author} {\bibfnamefont {J.}~\bibnamefont {Preskill}},\ }\bibfield  {title}
  {\bibinfo {title} {Topological quantum memory},\ }\href
  {https://doi.org/10.1063/1.1499754} {\bibfield  {journal} {\bibinfo
  {journal} {Journal of Mathematical Physics}\ }\textbf {\bibinfo {volume}
  {43}},\ \bibinfo {pages} {4452} (\bibinfo {year} {2002})}\BibitemShut
  {NoStop}%
\bibitem [{\citenamefont {Nishimori}(1981)}]{Nishimori1981}%
  \BibitemOpen
  \bibfield  {author} {\bibinfo {author} {\bibfnamefont {H.}~\bibnamefont
  {Nishimori}},\ }\bibfield  {title} {\bibinfo {title} {{Internal Energy,
  Specific Heat and Correlation Function of the Bond-Random Ising Model}},\
  }\href {https://doi.org/10.1143/PTP.66.1169} {\bibfield  {journal} {\bibinfo
  {journal} {Progress of Theoretical Physics}\ }\textbf {\bibinfo {volume}
  {66}},\ \bibinfo {pages} {1169} (\bibinfo {year} {1981})}\BibitemShut
  {NoStop}%
\bibitem [{\citenamefont {Nishimori}(1993)}]{Nishimori1993decoding}%
  \BibitemOpen
  \bibfield  {author} {\bibinfo {author} {\bibfnamefont {H.}~\bibnamefont
  {Nishimori}},\ }\bibfield  {title} {\bibinfo {title} {Optimum decoding
  temperature for error-correcting codes},\ }\href
  {https://doi.org/10.1143/JPSJ.62.2973} {\bibfield  {journal} {\bibinfo
  {journal} {Journal of the Physical Society of Japan}\ }\textbf {\bibinfo
  {volume} {62}},\ \bibinfo {pages} {2973} (\bibinfo {year}
  {1993})}\BibitemShut {NoStop}%
\bibitem [{\citenamefont {Garratt}\ \emph {et~al.}(2022)\citenamefont
  {Garratt}, \citenamefont {Weinstein},\ and\ \citenamefont
  {Altman}}]{Garratt22}%
  \BibitemOpen
  \bibfield  {author} {\bibinfo {author} {\bibfnamefont {S.~J.}\ \bibnamefont
  {Garratt}}, \bibinfo {author} {\bibfnamefont {Z.}~\bibnamefont {Weinstein}},\
  and\ \bibinfo {author} {\bibfnamefont {E.}~\bibnamefont {Altman}},\
  }\bibfield  {title} {\bibinfo {title} {{Measurements conspire nonlocally to
  restructure critical quantum states}},\ }\href@noop {} {\  (\bibinfo {year}
  {2022})},\ \Eprint {https://arxiv.org/abs/2207.09476} {arXiv:2207.09476}
  \BibitemShut {NoStop}%
\bibitem [{\citenamefont {Lee}\ \emph {et~al.}(2023)\citenamefont {Lee},
  \citenamefont {You},\ and\ \citenamefont {Xu}}]{lee2023symmetry}%
  \BibitemOpen
  \bibfield  {author} {\bibinfo {author} {\bibfnamefont {J.~Y.}\ \bibnamefont
  {Lee}}, \bibinfo {author} {\bibfnamefont {Y.-Z.}\ \bibnamefont {You}},\ and\
  \bibinfo {author} {\bibfnamefont {C.}~\bibnamefont {Xu}},\ }\href@noop {}
  {\bibinfo {title} {Symmetry protected topological phases under decoherence}}
  (\bibinfo {year} {2023}),\ \Eprint {https://arxiv.org/abs/2210.16323}
  {arXiv:2210.16323 [cond-mat.str-el]} \BibitemShut {NoStop}%
\bibitem [{\citenamefont {Chow}\ \emph {et~al.}(2011)\citenamefont {Chow},
  \citenamefont {C\'orcoles}, \citenamefont {Gambetta}, \citenamefont
  {Rigetti}, \citenamefont {Johnson}, \citenamefont {Smolin}, \citenamefont
  {Rozen}, \citenamefont {Keefe}, \citenamefont {Rothwell}, \citenamefont
  {Ketchen},\ and\ \citenamefont {Steffen}}]{chow2011simple}%
  \BibitemOpen
  \bibfield  {author} {\bibinfo {author} {\bibfnamefont {J.~M.}\ \bibnamefont
  {Chow}}, \bibinfo {author} {\bibfnamefont {A.~D.}\ \bibnamefont
  {C\'orcoles}}, \bibinfo {author} {\bibfnamefont {J.~M.}\ \bibnamefont
  {Gambetta}}, \bibinfo {author} {\bibfnamefont {C.}~\bibnamefont {Rigetti}},
  \bibinfo {author} {\bibfnamefont {B.~R.}\ \bibnamefont {Johnson}}, \bibinfo
  {author} {\bibfnamefont {J.~A.}\ \bibnamefont {Smolin}}, \bibinfo {author}
  {\bibfnamefont {J.~R.}\ \bibnamefont {Rozen}}, \bibinfo {author}
  {\bibfnamefont {G.~A.}\ \bibnamefont {Keefe}}, \bibinfo {author}
  {\bibfnamefont {M.~B.}\ \bibnamefont {Rothwell}}, \bibinfo {author}
  {\bibfnamefont {M.~B.}\ \bibnamefont {Ketchen}},\ and\ \bibinfo {author}
  {\bibfnamefont {M.}~\bibnamefont {Steffen}},\ }\bibfield  {title} {\bibinfo
  {title} {{Simple All-Microwave Entangling Gate for Fixed-Frequency
  Superconducting Qubits}},\ }\href
  {https://doi.org/10.1103/PhysRevLett.107.080502} {\bibfield  {journal}
  {\bibinfo  {journal} {Phys. Rev. Lett.}\ }\textbf {\bibinfo {volume} {107}},\
  \bibinfo {pages} {080502} (\bibinfo {year} {2011})}\BibitemShut {NoStop}%
\bibitem [{\citenamefont {Sheldon}\ \emph {et~al.}(2016)\citenamefont
  {Sheldon}, \citenamefont {Magesan}, \citenamefont {Chow},\ and\ \citenamefont
  {Gambetta}}]{sheldon2016ecr}%
  \BibitemOpen
  \bibfield  {author} {\bibinfo {author} {\bibfnamefont {S.}~\bibnamefont
  {Sheldon}}, \bibinfo {author} {\bibfnamefont {E.}~\bibnamefont {Magesan}},
  \bibinfo {author} {\bibfnamefont {J.~M.}\ \bibnamefont {Chow}},\ and\
  \bibinfo {author} {\bibfnamefont {J.~M.}\ \bibnamefont {Gambetta}},\
  }\bibfield  {title} {\bibinfo {title} {Procedure for systematically tuning up
  cross-talk in the cross-resonance gate},\ }\href
  {https://doi.org/10.1103/PhysRevA.93.060302} {\bibfield  {journal} {\bibinfo
  {journal} {Phys. Rev. A}\ }\textbf {\bibinfo {volume} {93}},\ \bibinfo
  {pages} {060302} (\bibinfo {year} {2016})}\BibitemShut {NoStop}%
\bibitem [{\citenamefont {Malekakhlagh}\ \emph {et~al.}(2020)\citenamefont
  {Malekakhlagh}, \citenamefont {Magesan},\ and\ \citenamefont
  {McKay}}]{malek2020ecr}%
  \BibitemOpen
  \bibfield  {author} {\bibinfo {author} {\bibfnamefont {M.}~\bibnamefont
  {Malekakhlagh}}, \bibinfo {author} {\bibfnamefont {E.}~\bibnamefont
  {Magesan}},\ and\ \bibinfo {author} {\bibfnamefont {D.~C.}\ \bibnamefont
  {McKay}},\ }\bibfield  {title} {\bibinfo {title} {First-principles analysis
  of cross-resonance gate operation},\ }\href
  {https://doi.org/10.1103/PhysRevA.102.042605} {\bibfield  {journal} {\bibinfo
   {journal} {Phys. Rev. A}\ }\textbf {\bibinfo {volume} {102}},\ \bibinfo
  {pages} {042605} (\bibinfo {year} {2020})}\BibitemShut {NoStop}%
\bibitem [{\citenamefont {Sundaresan}\ \emph {et~al.}(2020)\citenamefont
  {Sundaresan}, \citenamefont {Lauer}, \citenamefont {Pritchett}, \citenamefont
  {Magesan}, \citenamefont {Jurcevic},\ and\ \citenamefont
  {Gambetta}}]{sundaresan2020ecr}%
  \BibitemOpen
  \bibfield  {author} {\bibinfo {author} {\bibfnamefont {N.}~\bibnamefont
  {Sundaresan}}, \bibinfo {author} {\bibfnamefont {I.}~\bibnamefont {Lauer}},
  \bibinfo {author} {\bibfnamefont {E.}~\bibnamefont {Pritchett}}, \bibinfo
  {author} {\bibfnamefont {E.}~\bibnamefont {Magesan}}, \bibinfo {author}
  {\bibfnamefont {P.}~\bibnamefont {Jurcevic}},\ and\ \bibinfo {author}
  {\bibfnamefont {J.~M.}\ \bibnamefont {Gambetta}},\ }\bibfield  {title}
  {\bibinfo {title} {Reducing unitary and spectator errors in cross resonance
  with optimized rotary echoes},\ }\href
  {https://doi.org/10.1103/PRXQuantum.1.020318} {\bibfield  {journal} {\bibinfo
   {journal} {PRX Quantum}\ }\textbf {\bibinfo {volume} {1}},\ \bibinfo {pages}
  {020318} (\bibinfo {year} {2020})}\BibitemShut {NoStop}%
\bibitem [{Note1()}]{Note1}%
  \BibitemOpen
  \bibinfo {note} {IBM Quantum. \protect \href
  {https://quantum-computing.ibm.com/}{https://quantum-computing.ibm.com/},
  2021. (Downloaded August 6, 2023).}\BibitemShut {Stop}%
\bibitem [{sup()}]{supplement}%
  \BibitemOpen
  \href@noop {} {\bibinfo {title} {{See Supplemental Material.}}}\BibitemShut
  {Stop}%
\bibitem [{\citenamefont {Flammia}\ and\ \citenamefont
  {Liu}(2011)}]{Flammia11fidelity}%
  \BibitemOpen
  \bibfield  {author} {\bibinfo {author} {\bibfnamefont {S.~T.}\ \bibnamefont
  {Flammia}}\ and\ \bibinfo {author} {\bibfnamefont {Y.-K.}\ \bibnamefont
  {Liu}},\ }\bibfield  {title} {\bibinfo {title} {Direct fidelity estimation
  from few pauli measurements},\ }\href
  {https://doi.org/10.1103/PhysRevLett.106.230501} {\bibfield  {journal}
  {\bibinfo  {journal} {Phys. Rev. Lett.}\ }\textbf {\bibinfo {volume} {106}},\
  \bibinfo {pages} {230501} (\bibinfo {year} {2011})}\BibitemShut {NoStop}%
\bibitem [{\citenamefont {da~Silva}\ \emph {et~al.}(2011)\citenamefont
  {da~Silva}, \citenamefont {Landon-Cardinal},\ and\ \citenamefont
  {Poulin}}]{Silva11fidelity}%
  \BibitemOpen
  \bibfield  {author} {\bibinfo {author} {\bibfnamefont {M.~P.}\ \bibnamefont
  {da~Silva}}, \bibinfo {author} {\bibfnamefont {O.}~\bibnamefont
  {Landon-Cardinal}},\ and\ \bibinfo {author} {\bibfnamefont {D.}~\bibnamefont
  {Poulin}},\ }\bibfield  {title} {\bibinfo {title} {Practical characterization
  of quantum devices without tomography},\ }\href
  {https://doi.org/10.1103/PhysRevLett.107.210404} {\bibfield  {journal}
  {\bibinfo  {journal} {Phys. Rev. Lett.}\ }\textbf {\bibinfo {volume} {107}},\
  \bibinfo {pages} {210404} (\bibinfo {year} {2011})}\BibitemShut {NoStop}%
\bibitem [{\citenamefont {{Higgott}}()}]{PyMatching21}%
  \BibitemOpen
  \bibfield  {author} {\bibinfo {author} {\bibfnamefont {O.}~\bibnamefont
  {{Higgott}}},\ }\bibfield  {title} {\bibinfo {title} {{PyMatching: A Python
  package for decoding quantum codes with minimum-weight perfect matching}},\
  }\href@noop {} {\ }\Eprint {https://arxiv.org/abs/2105.13082}
  {arXiv:2105.13082} \BibitemShut {NoStop}%
\bibitem [{\citenamefont {de~Queiroz}(2006)}]{Queiroz2006exponents}%
  \BibitemOpen
  \bibfield  {author} {\bibinfo {author} {\bibfnamefont {S.~L.~A.}\
  \bibnamefont {de~Queiroz}},\ }\bibfield  {title} {\bibinfo {title}
  {{Multicritical point of Ising spin glasses on triangular and honeycomb
  lattices}},\ }\href {https://doi.org/10.1103/PhysRevB.73.064410} {\bibfield
  {journal} {\bibinfo  {journal} {Phys. Rev. B}\ }\textbf {\bibinfo {volume}
  {73}},\ \bibinfo {pages} {064410} (\bibinfo {year} {2006})}\BibitemShut
  {NoStop}%
\bibitem [{Note2()}]{Note2}%
  \BibitemOpen
  \bibinfo {note} {See Extended Data Fig.~\ref {fig:1Dv2D}}\BibitemShut
  {NoStop}%
\bibitem [{\citenamefont {Binder}\ and\ \citenamefont
  {Young}(1986)}]{Binder86rmp}%
  \BibitemOpen
  \bibfield  {author} {\bibinfo {author} {\bibfnamefont {K.}~\bibnamefont
  {Binder}}\ and\ \bibinfo {author} {\bibfnamefont {A.~P.}\ \bibnamefont
  {Young}},\ }\bibfield  {title} {\bibinfo {title} {{Spin glasses: Experimental
  facts, theoretical concepts, and open questions}},\ }\href
  {https://doi.org/10.1103/RevModPhys.58.801} {\bibfield  {journal} {\bibinfo
  {journal} {Rev. Mod. Phys.}\ }\textbf {\bibinfo {volume} {58}},\ \bibinfo
  {pages} {801} (\bibinfo {year} {1986})}\BibitemShut {NoStop}%
\bibitem [{\citenamefont {Iqbal}\ \emph
  {et~al.}(2023{\natexlab{a}})\citenamefont {Iqbal}, \citenamefont
  {Tantivasadakarn}, \citenamefont {Gatterman}, \citenamefont {Gerber},
  \citenamefont {Gilmore}, \citenamefont {Gresh}, \citenamefont {Hankin},
  \citenamefont {Hewitt}, \citenamefont {Horst}, \citenamefont {Matheny},
  \citenamefont {Mengle}, \citenamefont {Neyenhuis}, \citenamefont
  {Vishwanath}, \citenamefont {Foss-Feig}, \citenamefont {Verresen},\ and\
  \citenamefont {Dreyer}}]{iqbal2023topological}%
  \BibitemOpen
  \bibfield  {author} {\bibinfo {author} {\bibfnamefont {M.}~\bibnamefont
  {Iqbal}}, \bibinfo {author} {\bibfnamefont {N.}~\bibnamefont
  {Tantivasadakarn}}, \bibinfo {author} {\bibfnamefont {T.~M.}\ \bibnamefont
  {Gatterman}}, \bibinfo {author} {\bibfnamefont {J.~A.}\ \bibnamefont
  {Gerber}}, \bibinfo {author} {\bibfnamefont {K.}~\bibnamefont {Gilmore}},
  \bibinfo {author} {\bibfnamefont {D.}~\bibnamefont {Gresh}}, \bibinfo
  {author} {\bibfnamefont {A.}~\bibnamefont {Hankin}}, \bibinfo {author}
  {\bibfnamefont {N.}~\bibnamefont {Hewitt}}, \bibinfo {author} {\bibfnamefont
  {C.~V.}\ \bibnamefont {Horst}}, \bibinfo {author} {\bibfnamefont
  {M.}~\bibnamefont {Matheny}}, \bibinfo {author} {\bibfnamefont
  {T.}~\bibnamefont {Mengle}}, \bibinfo {author} {\bibfnamefont
  {B.}~\bibnamefont {Neyenhuis}}, \bibinfo {author} {\bibfnamefont
  {A.}~\bibnamefont {Vishwanath}}, \bibinfo {author} {\bibfnamefont
  {M.}~\bibnamefont {Foss-Feig}}, \bibinfo {author} {\bibfnamefont
  {R.}~\bibnamefont {Verresen}},\ and\ \bibinfo {author} {\bibfnamefont
  {H.}~\bibnamefont {Dreyer}},\ }\href@noop {} {\bibinfo {title} {{Topological
  Order from Measurements and Feed-Forward on a Trapped Ion Quantum Computer}}}
  (\bibinfo {year} {2023}{\natexlab{a}}),\ \Eprint
  {https://arxiv.org/abs/2302.01917} {arXiv:2302.01917} \BibitemShut {NoStop}%
\bibitem [{\citenamefont {Foss-Feig}\ \emph {et~al.}(2023)\citenamefont
  {Foss-Feig}, \citenamefont {Tikku}, \citenamefont {Lu}, \citenamefont
  {Mayer}, \citenamefont {Iqbal}, \citenamefont {Gatterman}, \citenamefont
  {Gerber}, \citenamefont {Gilmore}, \citenamefont {Gresh}, \citenamefont
  {Hankin}, \citenamefont {Hewitt}, \citenamefont {Horst}, \citenamefont
  {Matheny}, \citenamefont {Mengle}, \citenamefont {Neyenhuis}, \citenamefont
  {Dreyer}, \citenamefont {Hayes}, \citenamefont {Hsieh},\ and\ \citenamefont
  {Kim}}]{fossfeig2023experimental}%
  \BibitemOpen
  \bibfield  {author} {\bibinfo {author} {\bibfnamefont {M.}~\bibnamefont
  {Foss-Feig}}, \bibinfo {author} {\bibfnamefont {A.}~\bibnamefont {Tikku}},
  \bibinfo {author} {\bibfnamefont {T.-C.}\ \bibnamefont {Lu}}, \bibinfo
  {author} {\bibfnamefont {K.}~\bibnamefont {Mayer}}, \bibinfo {author}
  {\bibfnamefont {M.}~\bibnamefont {Iqbal}}, \bibinfo {author} {\bibfnamefont
  {T.~M.}\ \bibnamefont {Gatterman}}, \bibinfo {author} {\bibfnamefont {J.~A.}\
  \bibnamefont {Gerber}}, \bibinfo {author} {\bibfnamefont {K.}~\bibnamefont
  {Gilmore}}, \bibinfo {author} {\bibfnamefont {D.}~\bibnamefont {Gresh}},
  \bibinfo {author} {\bibfnamefont {A.}~\bibnamefont {Hankin}}, \bibinfo
  {author} {\bibfnamefont {N.}~\bibnamefont {Hewitt}}, \bibinfo {author}
  {\bibfnamefont {C.~V.}\ \bibnamefont {Horst}}, \bibinfo {author}
  {\bibfnamefont {M.}~\bibnamefont {Matheny}}, \bibinfo {author} {\bibfnamefont
  {T.}~\bibnamefont {Mengle}}, \bibinfo {author} {\bibfnamefont
  {B.}~\bibnamefont {Neyenhuis}}, \bibinfo {author} {\bibfnamefont
  {H.}~\bibnamefont {Dreyer}}, \bibinfo {author} {\bibfnamefont
  {D.}~\bibnamefont {Hayes}}, \bibinfo {author} {\bibfnamefont {T.~H.}\
  \bibnamefont {Hsieh}},\ and\ \bibinfo {author} {\bibfnamefont {I.~H.}\
  \bibnamefont {Kim}},\ }\href@noop {} {\bibinfo {title} {{Experimental
  demonstration of the advantage of adaptive quantum circuits}}} (\bibinfo
  {year} {2023}),\ \Eprint {https://arxiv.org/abs/2302.03029}
  {arXiv:2302.03029} \BibitemShut {NoStop}%
\bibitem [{\citenamefont {Iqbal}\ \emph
  {et~al.}(2023{\natexlab{b}})\citenamefont {Iqbal}, \citenamefont
  {Tantivasadakarn}, \citenamefont {Verresen}, \citenamefont {Campbell},
  \citenamefont {Dreiling}, \citenamefont {Figgatt}, \citenamefont {Gaebler},
  \citenamefont {Johansen}, \citenamefont {Mills}, \citenamefont {Moses},
  \citenamefont {Pino}, \citenamefont {Ransford}, \citenamefont {Rowe},
  \citenamefont {Siegfried}, \citenamefont {Stutz}, \citenamefont {Foss-Feig},
  \citenamefont {Vishwanath},\ and\ \citenamefont
  {Dreyer}}]{iqbal2023creation}%
  \BibitemOpen
  \bibfield  {author} {\bibinfo {author} {\bibfnamefont {M.}~\bibnamefont
  {Iqbal}}, \bibinfo {author} {\bibfnamefont {N.}~\bibnamefont
  {Tantivasadakarn}}, \bibinfo {author} {\bibfnamefont {R.}~\bibnamefont
  {Verresen}}, \bibinfo {author} {\bibfnamefont {S.~L.}\ \bibnamefont
  {Campbell}}, \bibinfo {author} {\bibfnamefont {J.~M.}\ \bibnamefont
  {Dreiling}}, \bibinfo {author} {\bibfnamefont {C.}~\bibnamefont {Figgatt}},
  \bibinfo {author} {\bibfnamefont {J.~P.}\ \bibnamefont {Gaebler}}, \bibinfo
  {author} {\bibfnamefont {J.}~\bibnamefont {Johansen}}, \bibinfo {author}
  {\bibfnamefont {M.}~\bibnamefont {Mills}}, \bibinfo {author} {\bibfnamefont
  {S.~A.}\ \bibnamefont {Moses}}, \bibinfo {author} {\bibfnamefont {J.~M.}\
  \bibnamefont {Pino}}, \bibinfo {author} {\bibfnamefont {A.}~\bibnamefont
  {Ransford}}, \bibinfo {author} {\bibfnamefont {M.}~\bibnamefont {Rowe}},
  \bibinfo {author} {\bibfnamefont {P.}~\bibnamefont {Siegfried}}, \bibinfo
  {author} {\bibfnamefont {R.~P.}\ \bibnamefont {Stutz}}, \bibinfo {author}
  {\bibfnamefont {M.}~\bibnamefont {Foss-Feig}}, \bibinfo {author}
  {\bibfnamefont {A.}~\bibnamefont {Vishwanath}},\ and\ \bibinfo {author}
  {\bibfnamefont {H.}~\bibnamefont {Dreyer}},\ }\href@noop {} {\bibinfo {title}
  {{Creation of Non-Abelian Topological Order and Anyons on a Trapped-Ion
  Processor}}} (\bibinfo {year} {2023}{\natexlab{b}}),\ \Eprint
  {https://arxiv.org/abs/2305.03766} {arXiv:2305.03766} \BibitemShut {NoStop}%
\bibitem [{\citenamefont {B{\"a}umer}\ \emph {et~al.}(2023)\citenamefont
  {B{\"a}umer}, \citenamefont {Tripathi}, \citenamefont {Wang}, \citenamefont
  {Rall}, \citenamefont {Chen}, \citenamefont {Majumder}, \citenamefont
  {Seif},\ and\ \citenamefont {Minev}}]{baumer2023efficient}%
  \BibitemOpen
  \bibfield  {author} {\bibinfo {author} {\bibfnamefont {E.}~\bibnamefont
  {B{\"a}umer}}, \bibinfo {author} {\bibfnamefont {V.}~\bibnamefont
  {Tripathi}}, \bibinfo {author} {\bibfnamefont {D.~S.}\ \bibnamefont {Wang}},
  \bibinfo {author} {\bibfnamefont {P.}~\bibnamefont {Rall}}, \bibinfo {author}
  {\bibfnamefont {E.~H.}\ \bibnamefont {Chen}}, \bibinfo {author}
  {\bibfnamefont {S.}~\bibnamefont {Majumder}}, \bibinfo {author}
  {\bibfnamefont {A.}~\bibnamefont {Seif}},\ and\ \bibinfo {author}
  {\bibfnamefont {Z.~K.}\ \bibnamefont {Minev}},\ }\bibfield  {title} {\bibinfo
  {title} {Efficient long-range entanglement using dynamic circuits},\
  }\href@noop {} {\bibfield  {journal} {\bibinfo  {journal} {arXiv preprint
  arXiv:2308.13065}\ } (\bibinfo {year} {2023})},\ \Eprint
  {https://arxiv.org/abs/arXiv:2308.13065} {arXiv:2308.13065} \BibitemShut
  {NoStop}%
\bibitem [{\citenamefont {Hastings}\ and\ \citenamefont
  {Haah}(2021)}]{Hastings2021dynamically}%
  \BibitemOpen
  \bibfield  {author} {\bibinfo {author} {\bibfnamefont {M.~B.}\ \bibnamefont
  {Hastings}}\ and\ \bibinfo {author} {\bibfnamefont {J.}~\bibnamefont
  {Haah}},\ }\bibfield  {title} {\bibinfo {title} {Dynamically {G}enerated
  {L}ogical {Q}ubits},\ }\href {https://doi.org/10.22331/q-2021-10-19-564}
  {\bibfield  {journal} {\bibinfo  {journal} {{Quantum}}\ }\textbf {\bibinfo
  {volume} {5}},\ \bibinfo {pages} {564} (\bibinfo {year} {2021})}\BibitemShut
  {NoStop}%
\bibitem [{\citenamefont {Kim}\ \emph {et~al.}(2023{\natexlab{a}})\citenamefont
  {Kim}, \citenamefont {Eddins}, \citenamefont {Anand}, \citenamefont {Wei},
  \citenamefont {Van Den~Berg}, \citenamefont {Rosenblatt}, \citenamefont
  {Nayfeh}, \citenamefont {Wu}, \citenamefont {Zaletel}, \citenamefont {Temme}
  \emph {et~al.}}]{kim2023evidence}%
  \BibitemOpen
  \bibfield  {author} {\bibinfo {author} {\bibfnamefont {Y.}~\bibnamefont
  {Kim}}, \bibinfo {author} {\bibfnamefont {A.}~\bibnamefont {Eddins}},
  \bibinfo {author} {\bibfnamefont {S.}~\bibnamefont {Anand}}, \bibinfo
  {author} {\bibfnamefont {K.~X.}\ \bibnamefont {Wei}}, \bibinfo {author}
  {\bibfnamefont {E.}~\bibnamefont {Van Den~Berg}}, \bibinfo {author}
  {\bibfnamefont {S.}~\bibnamefont {Rosenblatt}}, \bibinfo {author}
  {\bibfnamefont {H.}~\bibnamefont {Nayfeh}}, \bibinfo {author} {\bibfnamefont
  {Y.}~\bibnamefont {Wu}}, \bibinfo {author} {\bibfnamefont {M.}~\bibnamefont
  {Zaletel}}, \bibinfo {author} {\bibfnamefont {K.}~\bibnamefont {Temme}},
  \emph {et~al.},\ }\bibfield  {title} {\bibinfo {title} {Evidence for the
  utility of quantum computing before fault tolerance},\ }\href
  {https://doi.org/10.1038/s41586-023-06096-3} {\bibfield  {journal} {\bibinfo
  {journal} {Nature}\ }\textbf {\bibinfo {volume} {618}},\ \bibinfo {pages}
  {500} (\bibinfo {year} {2023}{\natexlab{a}})}\BibitemShut {NoStop}%
\bibitem [{\citenamefont {Earnest}\ \emph {et~al.}(2021)\citenamefont
  {Earnest}, \citenamefont {Tornow},\ and\ \citenamefont
  {Egger}}]{earnest2021pulse}%
  \BibitemOpen
  \bibfield  {author} {\bibinfo {author} {\bibfnamefont {N.}~\bibnamefont
  {Earnest}}, \bibinfo {author} {\bibfnamefont {C.}~\bibnamefont {Tornow}},\
  and\ \bibinfo {author} {\bibfnamefont {D.~J.}\ \bibnamefont {Egger}},\
  }\bibfield  {title} {\bibinfo {title} {{Pulse-efficient circuit transpilation
  for quantum applications on cross-resonance-based hardware}},\ }\href
  {https://doi.org/10.1103/PhysRevResearch.3.043088} {\bibfield  {journal}
  {\bibinfo  {journal} {Phys. Rev. Res.}\ }\textbf {\bibinfo {volume} {3}},\
  \bibinfo {pages} {043088} (\bibinfo {year} {2021})}\BibitemShut {NoStop}%
\bibitem [{\citenamefont {Kim}\ \emph {et~al.}(2023{\natexlab{b}})\citenamefont
  {Kim}, \citenamefont {Wood}, \citenamefont {Yoder}, \citenamefont {Merkel},
  \citenamefont {Gambetta}, \citenamefont {Temme},\ and\ \citenamefont
  {Kandala}}]{kim2023scalable}%
  \BibitemOpen
  \bibfield  {author} {\bibinfo {author} {\bibfnamefont {Y.}~\bibnamefont
  {Kim}}, \bibinfo {author} {\bibfnamefont {C.~J.}\ \bibnamefont {Wood}},
  \bibinfo {author} {\bibfnamefont {T.~J.}\ \bibnamefont {Yoder}}, \bibinfo
  {author} {\bibfnamefont {S.~T.}\ \bibnamefont {Merkel}}, \bibinfo {author}
  {\bibfnamefont {J.~M.}\ \bibnamefont {Gambetta}}, \bibinfo {author}
  {\bibfnamefont {K.}~\bibnamefont {Temme}},\ and\ \bibinfo {author}
  {\bibfnamefont {A.}~\bibnamefont {Kandala}},\ }\bibfield  {title} {\bibinfo
  {title} {{Scalable error mitigation for noisy quantum circuits produces
  competitive expectation values}},\ }\href
  {https://doi.org/10.1038/s41567-022-01914-3} {\bibfield  {journal} {\bibinfo
  {journal} {Nature Physics}\ }\textbf {\bibinfo {volume} {19}},\ \bibinfo
  {pages} {752} (\bibinfo {year} {2023}{\natexlab{b}})}\BibitemShut {NoStop}%
\bibitem [{\citenamefont {van~den Berg}\ \emph {et~al.}(2022)\citenamefont
  {van~den Berg}, \citenamefont {Minev},\ and\ \citenamefont
  {Temme}}]{van2022model}%
  \BibitemOpen
  \bibfield  {author} {\bibinfo {author} {\bibfnamefont {E.}~\bibnamefont
  {van~den Berg}}, \bibinfo {author} {\bibfnamefont {Z.~K.}\ \bibnamefont
  {Minev}},\ and\ \bibinfo {author} {\bibfnamefont {K.}~\bibnamefont {Temme}},\
  }\bibfield  {title} {\bibinfo {title} {Model-free readout-error mitigation
  for quantum expectation values},\ }\href
  {https://doi.org/10.1103/PhysRevA.105.032620} {\bibfield  {journal} {\bibinfo
   {journal} {Phys. Rev. A}\ }\textbf {\bibinfo {volume} {105}},\ \bibinfo
  {pages} {032620} (\bibinfo {year} {2022})}\BibitemShut {NoStop}%
\end{thebibliography}%
\clearpage
\newpage
\section*{Extended data}

\begin{figure*}[h!]
\includegraphics[width=.9\columnwidth]{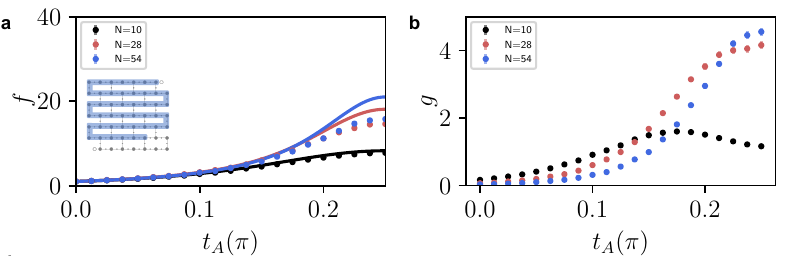}
\caption{
{\bf Absence of finite threshold in one-dimensional protocol}, for comparison with Fig.~\ref{fig:fig4}. 
(a) $f$ grows with increasing system size but converges to finite value that depends on $t_A$. 
(b) The peak of $g$ converges to $t_A=\pi/4$ indicative of absence of finite threshold for coherent error. 
}
\label{fig:1Dv2D}
\end{figure*}

\begin{figure*}[h!]
\includegraphics[width=.95\columnwidth]{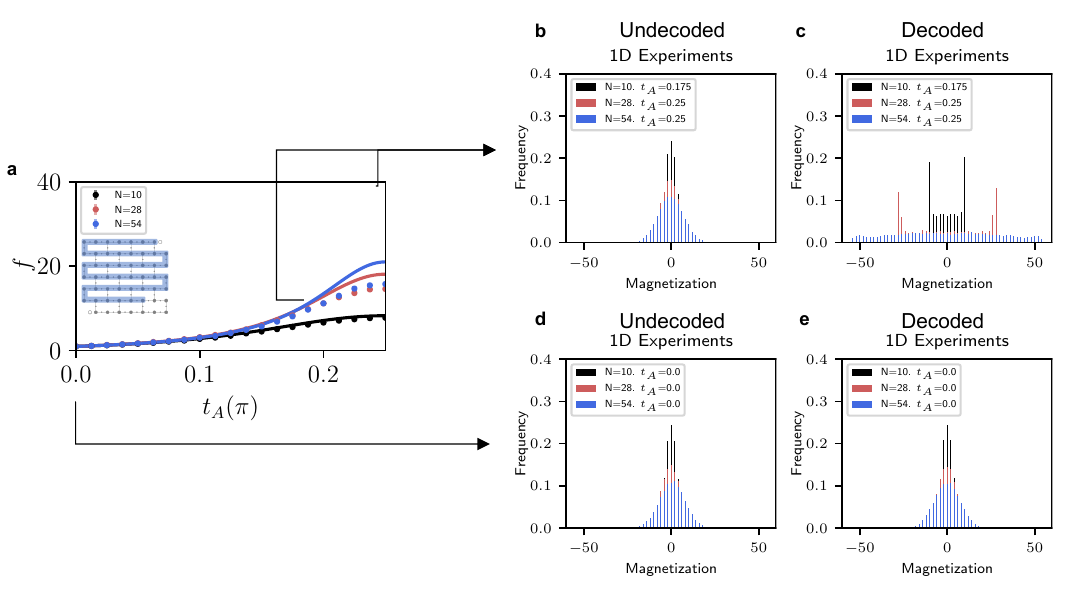}
\caption{\label{fig:fig3c1D}
\textbf{Magnetization of 1D experiments with and without decoding at different $t_A$ values} 
(\textbf{a}) Two-point correlations in 1D experiments for sweeps of $t_A$. The histograms at values of $t_A$ where variances were maximized for undecoded (\textbf{b}) and decoded (\textbf{c}). Although the bimodal distribution persisted up to a system size of 28, at 54 the distribution became uniform. And as expected, both the undecoded (\textbf{d}) and decoded (\textbf{e}) exhibited a binomial distribution in the trivial state ($t_A=0$). 
}
\end{figure*}

\begin{figure*}[h!]
\includegraphics[width=.95\columnwidth]{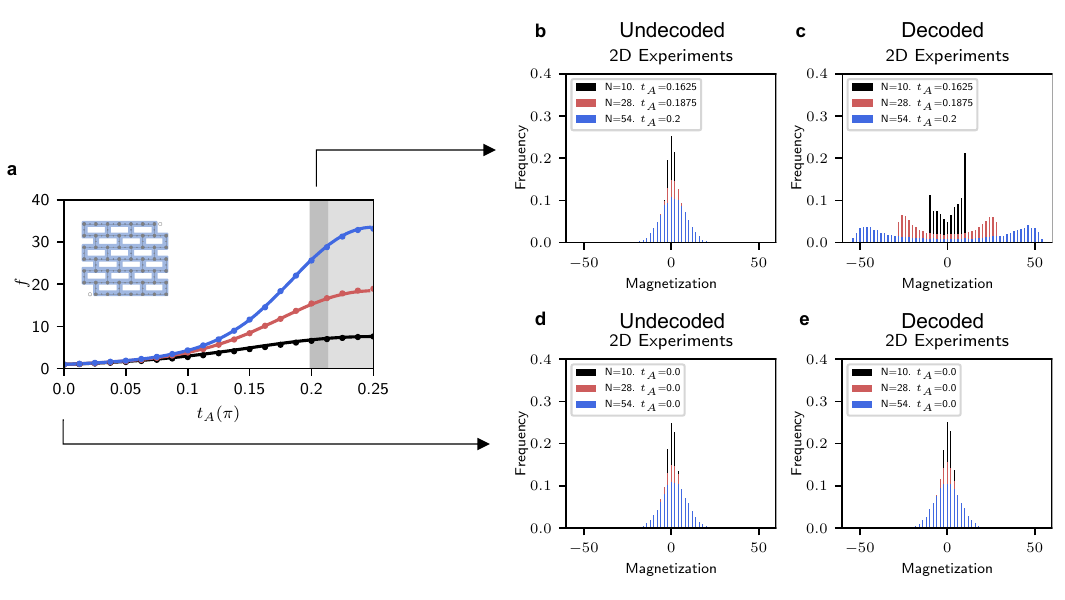}
\caption{\label{fig:fig3c2D}
\textbf{Magnetization of 2D experiments with and without decoding at different $t_A$ values} 
(\textbf{a}) Two-point correlations in 2D experiments for sweeps of $t_A$. The histograms at values of $t_A$ where variances were maximized for undecoded (\textbf{b}) and decoded (\textbf{c}). In contrast to the 1D cases (Fig.~\ref{fig:fig3c1D}), the bimodal distribution persisted up to a system size of 54. And similarly to the 1D case, both the undecoded (\textbf{d}) and decoded (\textbf{e}) exhibited a binomial distribution in the trivial state ($t_A=0$).
}
\end{figure*}

As discussed in the main text, the 2D protocol exhibited robustness over the 1D protocol; the key signature being based on the scaling of the average of two-point correlations, $f$, as a function of system size (Fig.~\ref{fig:1Dv2D}). Whereas in 2D (Fig.~\ref{fig:fig3c2D}), we not only observed the expected $f\propto L_y^{1.9}$ behavior in the long-range ordered state, but also that the criticality occurs below the GHZ point ($t_A^c < \pi/4$). The 1D behavior (Fig.~\ref{fig:fig3c1D}), in contrast, exhibited no growth in $f$ with system size from 28 to 54, and had peak variances at the GHZ value of $t_A=\pi/4$. 
The binomial distributions (approaching Gaussian in asymptotic large system sizes) of magnetization in Fig.~\ref{fig:fig3c1D}bd and Fig.~\ref{fig:fig3c2D}bd arise because all the possible system bit-strings are equally probable in the pre-measurement system~\cite{zhu_nishimoris_2022}, which maintains the same in the trivial state at Fig.~\ref{fig:fig3c2D}e. 
\\

\onecolumngrid
\vskip 1cm

\twocolumngrid

\end{document}